\documentclass[floatfix,twocolumn,showpacs,preprintnumbers,amsmath,amssymb,prb]{revtex4}

\usepackage{graphicx}
\begin{document}

\title{

First-Principles Study of Substitutional Metal Impurities in Graphene: 
Structural, Electronic and Magnetic Properties}

\author{E.~J.~G.~Santos}
\email{eltonjose_gomes@ehu.es}
\affiliation{
Centro de F\'{\i}sica de Materiales CFM-MPC,
Centro Mixto CSIC-UPV/EHU, Apdo. 1072,
20080 San Sebasti\'an, Spain}
\affiliation{Donostia International Physics Center (DIPC),
Paseo Manuel de Lardizabal 4, 20018 San Sebasti\'an, Spain}

\author{A. Ayuela}
\email{swxayfea@ehu.es}
\affiliation{
Centro de F\'{\i}sica de Materiales CFM-MPC,
Centro Mixto CSIC-UPV/EHU, Apdo. 1072,
20080 San Sebasti\'an, Spain}
\affiliation{Donostia International Physics Center (DIPC),
Paseo Manuel de Lardizabal 4, 20018 San Sebasti\'an, Spain}

\author{D. S\'anchez-Portal}
\email{sqbsapod@ehu.es}
\affiliation{
Centro de F\'{\i}sica de Materiales CFM-MPC,
Centro Mixto CSIC-UPV/EHU, Apdo. 1072,
20080 San Sebasti\'an, Spain}
\affiliation{Donostia International Physics Center (DIPC),
Paseo Manuel de Lardizabal 4, 20018 San Sebasti\'an, Spain}

\date{\today}

\begin{abstract}
We present a theoretical study using
density functional calculations of the structural,
electronic and magnetic properties of 3$d$ transition metal,
noble metal and Zn atoms interacting with carbon monovacancies in graphene.
We pay special attention to the electronic and magnetic properties
of these substitutional impurities and found that 
they can be fully
understood using a simple model based on the hybridization between
the states of the metal atom, particularly the $d$ shell, and
the defect levels associated with an unreconstructed D$_{3h}$ 
carbon vacancy.
We identify three different regimes
associated with  the occupation of 
different carbon-metal
hybridized electronic levels:
({\it i})
bonding states are completely filled for
Sc and Ti, and these impurities are non-magnetic;
({\it ii}) the non-bonding $d$ shell is partially occupied 
for V, Cr and Mn and, correspondingly, 
these impurties present large and localized spin moments;
({\it iii}) antibonding states with increasing carbon character
are progressively filled for Co, Ni, the
noble metals and Zn.
The spin moments of these impurities 
oscillate
between 0 and 1~$\mu_B$ and are increasingly delocalized.
The substitutional Zn suffers a Jahn-Teller-like
distortion
from the $C_{3v}$ symmetry and, as a consequence, has a 
zero spin moment.
Fe occupies a distinct position at the border
between regimes ({\it ii})
and ({\it iii}) and shows a more complex behavior: while is non-magnetic
at the level of GGA calculations, 
its spin moment can be switched on 
using GGA+U calculations with moderate values of the U parameter.
\end{abstract}

\pacs{ 73.22.-f, 73.20.Hb, 75.20.Hr, 61.48.De}

\maketitle

\section{Introduction}

The electronic properties of two-dimensional graphene 
are currently the subject of intense experimental and
theoretical research.~\cite{Geim07,Neto09}
Graphene exhibits many intriguing phenomena
stemming from 
the characteristic conical dispersion and chiral behavior
of its valence and 
conduction bands nearby the Fermi energy.~\cite{Novoselov05,Katsnelson06,Park08}
In general, graphenic
nanostructures
like graphene nanoribbons, carbon nanotubes and
their interconnections,  are opening new routes
for research in the field
of nanoelectronics~\cite{Avouris07}, due to
the long spin relaxation and
decoherence times of these 
materials~\cite{Trauzettel07}. In
addition,
the possibility to control 
the magnetism of edge states in
nanoribbons 
by applying
external electric fields,~\cite{Son06,Yazyev08} 
make them very attractive for spintronic 
applications.~\cite{Hueso07}
However, for the design of realistic devices the effect
of defects and impurities have to be taken into account.  
For this reason, quite a lot of work has been devoted
to the study of defects and different types of impurities in these
materials.
It has now become clear that defects and dopants severely
affect some the properties of graphenic systems and can be used
to tune their response.
For example, the strong interplay between the presence of defect or dopants 
and the magnetic properties of carbon nanostructures has been 
stressed by several 
authors.~\cite{Esquinazi03,Lehtinen03,Lehtinen04,Palacios06,Krasheninnikov07,Uchoa08,Yazyev08bis,Palacios08,Rossier07}

\begin{figure}
\includegraphics[width=3.250in]{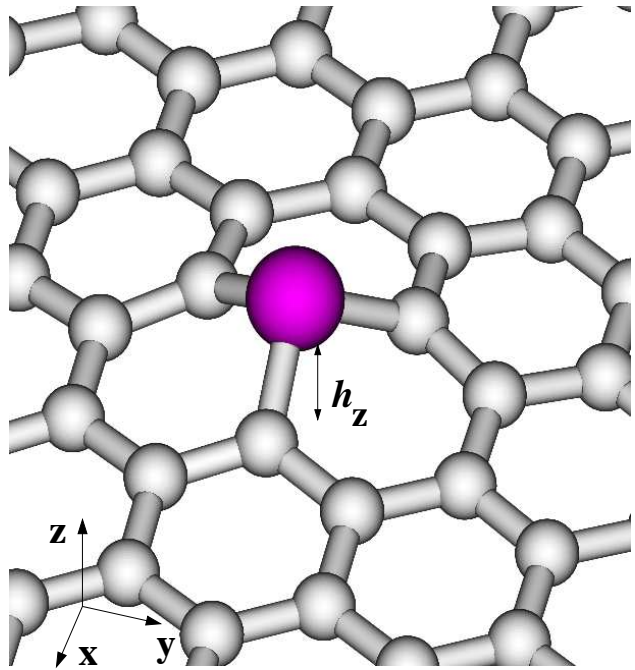}
\caption{\label{fig:fig1} (Color online) Typical geometry of 
transition and noble 
substitutional metal
atoms
in graphene. The metal atom moves
upwards
from the layer 
and occupies, in most cases, an almost perfectly symmetric 
three-fold position with C$_{3v}$ symmetry.
}
\end{figure}

Here we focus on substitutional impurities in graphene,
in which a single 
metal atom substitutes one or several carbon atoms in the layer.
Direct experimental evidence of the existence of these kind
of defects has been recently provided by Gan {\it et al.}~\cite{Banhart08}.
Using 
high-resolution transmission electron microscopy (HRTEM),
these authors were able to visualize individual
Au and Pt atoms incorporated into  a very thin graphitic layer probably
consisting of one or two graphene layers. From the
real-time evolution and temperature dependence of the 
dynamics they obtained information 
about the diffusion of these atoms. Large diffusion
barriers ($\sim$2.5~eV) were observed for in-plane migration, which
indicates the large
stability of these defects and the presence 
of strong carbon-metal bonds. These observations 
seem to indicated that the atoms occupy substitutional 
positions. 
There also exists evidence for the existence
of substitutional Ni impurities in
single-walled carbon nanotubes (SWCNT)~\cite{Ushiro06,Santos08}
and graphitic particles.~\cite{Banhart00}
Ushiro {\it et al.}~\cite{Ushiro06} showed that Ni
substitutional defects were present in  SWCNT samples even after
careful purification and that, according to their
analysis of the X-ray absorption data, the most
likely configuration was one in which the Ni atom
replaces a carbon atom, as seen in Fig. \ref{fig:fig1}.

The presence
of substitutional defects can have important implications
for the interpretation of some experimental 
evidence.
For example, 
substitutionals of magnetic transition metals
are expected to strongly influence the magnetic properties
of graphenic nanostructures.  Interestingly transition metals,
like Fe, Ni or Co, are among the most common catalysts
used for the production of SWCNT~\cite{growth}. Furthermore, 
Rodriguez-Manzo and Banhart~\cite{Banhart09} have recently
demonstrated the possibility to create individual vacancies at 
desired locations in carbon nanotubes using electron beams.
This ability, in combination with the observed stability of
substitutional impurities, can  open a route 
to fabricate new devices incorporating substitutional
impurities in certain locations or 
arranged in particular ways. Such devices would allow
for experimental
verification of some of the
unsual magnetic interactions mediated by the
graphenic carbon network that have been predicted
recently.~\cite{Brey07,Kirwan08,Santos09}

In this work we study theoretically
the structural,
electronic and magnetic properties of 3$d$ transition metals, 
noble metals and Zn as substitutional impurities in 
graphene using
density functional (DFT) calculations.
We only consider the configuration 
proposed by Ushiro {\it et al.},~\cite{Ushiro06}
in which a single metal atom
binds to a carbon monovacancy. Throughout the paper we will 
refer to this structure as substitutional configuration.
The results of our DFT calculations are in good agreement with the results
of other recent studies on
similar systems.~\cite{Santos08,Krasheninnikov09,Malola09,Boukhvalov09,Santos09,Ewels09}
In particular, Krasheninnikov {\it et al.}~\cite{Krasheninnikov09} have
presented the most complete first-principles
study to date of metal atoms interacting with single and
double vacancies in graphene. In most cases our results are in good
agreement with their predictions and similar general trends are found.
However, the electronic structure of these defects
has not been analyzed in detail to date and
no simple model to understand the observed behaviors has been
presented so far. Here we present the
evolution of the electronic structure of the substitutional defects
as we move along the transition series and correlate the observed
changes
with a simple picture of the metal-vacancy interaction.

In our study
we pay special attention to the electronic and magnetic properties
of the metal impurities. One of our key results is that the
electronic and magnetic properties of these 
substitutional metals can be fully
understood using a simple model based on the hybridization between
the states of the metal atom, particularly the $d$ shell, and
the defect levels associated with the unreconstructed carbon vacancy. 
The predictions of this model are in good agreement with
the calculated DFT band structures. With this model we can 
easily understand the
non-trivial behavior found for the binding energy and for the size
and localization of the spin moment
as we increase the number
of valence electrons in the impurity. 
In brief, we have identified three different regimes 
that can be correlated with the electron filling of
different carbon-metal 
hybridized levels:
({\it i}) 
bonding states are filled for
Sc and Ti, and these impurities are non-magnetic; 
({\it ii}) the non-bonding $d$ shell is partially occupied for V, Cr and 
Mn and, correspondingly, these 
impurities
present large and localized spin moments;
({\it iii}) antibonding states with increasing carbon character
are progressively filled for Co, Ni, the
noble metals and Zn, giving rise to  spin moments that oscillate
between 0 and 1~$\mu_B$ and are increasingly delocalized. 

We have also found that Zn becomes non-magnetic due 
to a Jahn-Teller distortion. However,
it is possible to stabilize a symmetric configuration with a spin moment
of 2$\mu_B$ with a very small energy penalty of $\sim$150~meV. Finally,
our calculations confirm that, the unexpected result that
Au substitutionals~\cite{Malola09,Krasheninnikov09} present a spin moment
of 1$\mu_B$, also holds for Ag and Cu, and thus stems only from the
number of valence electrons (see the Slater-Pauling-like plot
in Fig.~\ref{fig:fig3}). 

We have studied with special detail the complex case of Fe. 
This impurity occupies a distinct position
at the boundary between two different regimes and its
magnetic behavior
stems from the 
competition between the carbon-metal hybridization and 
the electron-electron interaction within the 3$d$ shell. As a result,
the spin moment of Fe is specially difficult to describe:
although non-magnetic using standard functionals within the 
generalized gradient approximation
(GGA)~\cite{gga},  the
magnetism of Fe 
appears 
using the so-called LDA+U methodology with reasonably low values of the
U parameter.

The paper is organized as follows. After a brief description of 
the computational approach in Sec.~\ref{sec:methods}, we present
a summary of the structure, energetics and magnetic properties
of all the studied elements in Sec.~\ref{sec:summary}. 
In this section
we also 
indicate
the general ideas behind our model of the 
metal-carbon hybrization in these systems. In Sec.~\ref{sec:D3hvac} the
electronic structure of the unreconstructed D$_{3h}$ 
carbon vacancy in graphene is
presented. This is one of the key ingredients to understand the binding
and electronic structure of substitutional impurities in graphene.
The electronic structure of the different groups of impurities
is described in the subsequent sections, particularly
in Sec.~\ref{sec:discussion}. 
The Zn substitutional impurity with its Janh-Teller distorsions is
described in Sec.~\ref{sec:Zn}.
A special section 
(Sec.~\ref{sec:FeMnborder}) is devoted to describe the special
role of Fe at the border between two different regimes. 
Finally, we close with some general conclusions.

\section{Theoretical Methods}
\label{sec:methods}

We have used two different approaches to perform
our DFT calculations: the
{\sc Siesta} method~\cite{siesta1,siesta2,siesta3}
using a basis set of localized 
numerical-atomic-orbitals~\cite{siesta2,junquera01} (NAOs) and the {\sc Vasp}
code\cite{vasp1,vasp2} using a basis set of plane waves.
We have used the Perdew-Burke-Ernzerhof~\cite{gga} 
GGA (PBE-GGA) functional in all our calculations. 
Most of our results are obtained
using a 4$\times$4 supercell. This supercell
is
sufficiently large to obtain reliable 
results. This 
has been proven by performing calculations 
using larger supercells up to 8$\times$8 for several elements.
The utilized codes perform three-dimensional periodic calculations.
In order to avoid spurious interactions between periodic images
of the defective graphene layer, the 
size of the supercell 
perpendicular
to the plane was always
larger than  15~\AA. The convergence with respect to the number
of $k$-points was
specially critical to obtain accurate results for the spin moment
in the studied systems. For all the impurities and different
supercell sizes we have used a large 
number of $k$-points,  
consistent with a 136$\times$136 Monkhorst-Pack~\cite{MonkhorstPack}
sampling of the unit cell of graphene, in combination 
with a Fermi-smearing of 21~meV.
All the atomic coordinates were always optimized
until forces along all directions were smaller than 0.05 eV/\AA.

For the 
{\sc Siesta} calculations we have used 
Troullier-Martins norm-conserving 
pseudopotentials~\cite{TM} generated using the pseudization radii 
shown 
in the Appendix.
The pseudopotentials for the metal atoms
include nonlinear core corrections~\cite{nlcc} for exchange
and correlation. The pseudocore 
radii ($r_{core}$) have been optimized
for each element and are also presented
in the Appedix.
Using
nonlinear core corrections is known to be critical 
to describe properly the spin moment
and magnetic properties of transition metals. 
We have tested that these pseudopotentials yield the correct
spin moments and  band structures in bulk phases.

The spacing of the real-space grid used to calculate the
Hartree and exchange-correlation contribution to the total
energy and Hamiltonian with {\sc Siesta} was equivalent to a 180 Ry
plane-wave cutoff.
A double-$\zeta$ polarized (DZP)~\cite{junquera01,siesta2} 
basis set has been used for the calculation of the magnetic 
and electronic properties.
However, we have checked that using a
double-$\zeta$ (DZ) basis set for carbon yields almost identical
relaxed structures as the DZP basis and, therefore, 
we have used the smaller DZ basis for most of the
structural relaxations. 
The shape of the basis orbitals was automatically determined
by {\sc Siesta} using the algorithms described in Ref.~\onlinecite{siesta2}. 
The cutoff radii of the different orbitals was obtained 
using an \textit{energy shift} of 50~meV. Although this basis set
proved to be sufficiently accurate to describe the geometries, spin moments
and band structures, for some metal atoms the binding energies where slightly 
overestimated. For those atoms the radii of the basis orbitals were enlarged
(using smaller values of the \textit{energy shift} parameter)
until binding energies were converged within a few tens of meV. The
resulting radii are shown 
in the Appendix.

With the {\sc Vasp} code we have used a well converged
plane-wave cutoff energy 
of 400 eV combined with the projected-augmented-wave (PAW)
method.
Using 
the provided
PAW potentials allows
to check the possible limitations of  pseudopotential
calculations. However, as shown below, the agreement
between both sets of calculations is excellent.
We have also performed GGA+U calculations using the 
formulation of Dudarev {\it et al.}~\cite{Dudarev98}.
 In this
formulation a single U parameter is used, which we have 
taken as an empirical parameter and 
varied in the range 1-4.5~eV in our calculations
for all the studied impurities. Only for Fe, GGA+U results
showed significant (qualitative) differences respect to those
of PBE-GGA calculations.

\begin{figure}
\includegraphics[width=3.250in]{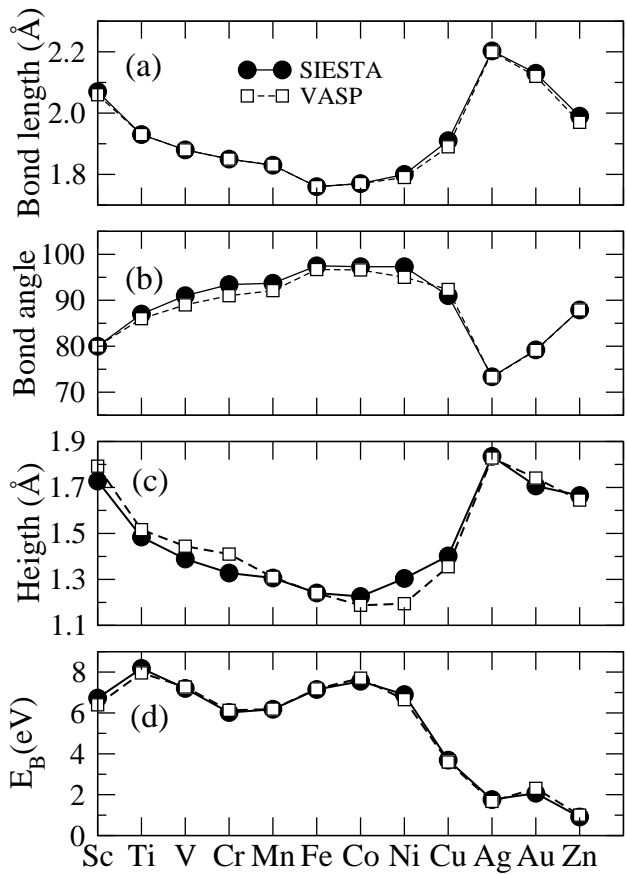}
\caption{\label{fig:fig2} 
Structural
parameters and 
binding energies of
substitutional transition and noble metals in graphene. 
Bond lengths and angles
have been averaged for the noble metals.
The data presented for Zn correspond to the high-spin
solution with C$_{3v}$ symmetry, and are very close to the
averaged results for the most stable distorted solution.
}
\end{figure}

\begin{figure}
\includegraphics[width=3.250in]{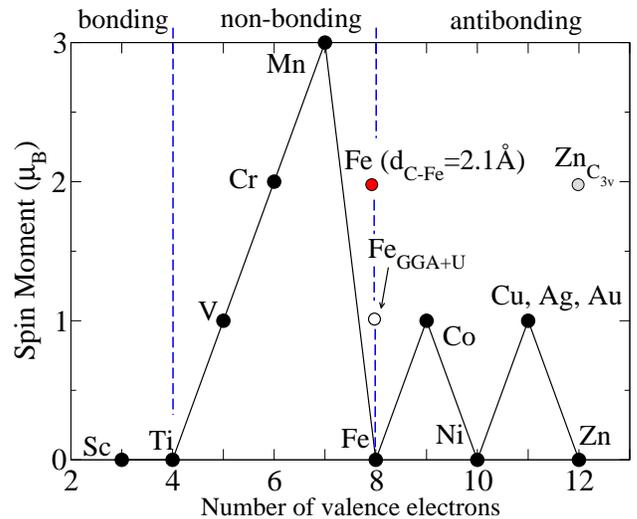}
\caption{\label{fig:fig3} (Color online) Spin moment of
substitutional transition and noble metals in graphene
as a function of the number of valence electrons (Slater-Pauling-type
plot). 
Black symbols correspond to the
most stable configurations using GGA. Results
are almost identical using {\sc Siesta} and {\sc Vasp} codes. Three
main regimes are found as explained in detail 
in the text: ({\it i}) filling of the metal-carbon bonding 
states gives rise to the non-magnetic behavior of Ti and Sc;
({\it ii}) non-bonding $d$ states are filled for V, Cr and
Mn giving rise to high spin moments; ({\it iii})
for Fe all non-bonding levels are occupied and 
metal-carbon antibonding states start to be filled
giving rise to the observed oscillatory behavior for
Co, Ni, Cu and Zn. 
Open and gray (red online) symbols correspond, respectively, to 
calculations of Fe using GGA+U and artificially increasing 
the height of the metal atom over the graphene layer (see the text).
Symbol marked
as Zn$_{{\text C}_{3v}}$ corresponds to a Zn impurity
in a high-spin symmetric C$_{3v}$ configuration.
}
\end{figure}

\begin{figure}
\includegraphics[width=3.250in]{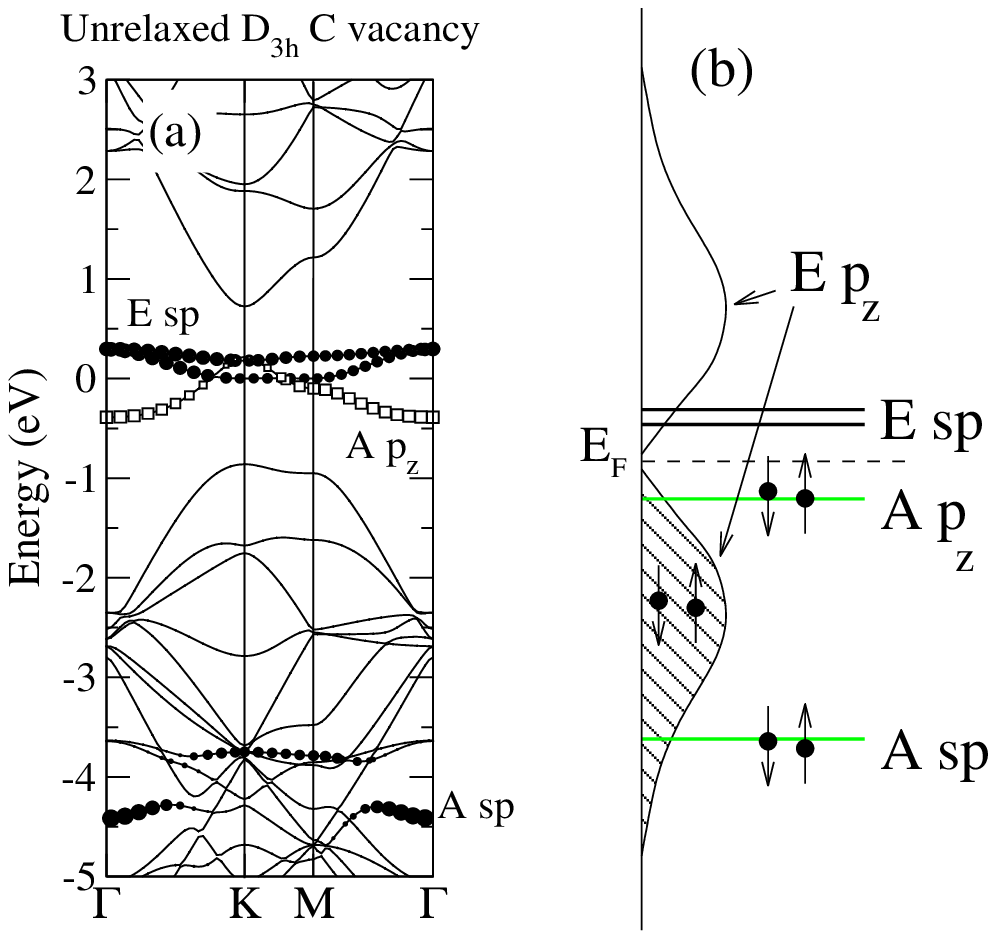}
\caption{\label{fig:fig4} (Color online) 
(a) Calculated band structure for an unrelaxed carbon 
vacancy (D$_{3h}$ symmetry) in a 4$\times$4 supercell
of graphene. Symbols indicated those
bands with larger weight on the carbon atoms around the 
vacancy (solid symbols for bands with $sp$ character
and open symbols for a band with $p_z$ character).
The electronic structure near E$_F$ is
dominated by a fully symmetric  $p_z$ level (A $p_z$) and two 
 defect levels with E symmetry and $sp$ character (E $sp$).
Notice that, due to the strong hybridization
with the rest of the graphene layer, it is not possible to 
identify well defined defect levels
with E symmetry and $p_z$ character.  
(b) Scheme of the electronic structure of the D$_{3h}$ C vacancy
indicating the character and symmetry of the different levels
and their occupations. Signal associated with the E $p_z$ level
extends over the whole valence and conduction band.}
\end{figure}

\section{ Main properties of substitutional transition metals in graphene}
\label{sec:summary}

We provide in this section a brief 
summary of our results for the structure, binding,
and spin moments of subtitutional 
3$d$ transition metals, noble metals and Zn
in graphene.

\subsection{Geometry and structural parameters}

The typical geometry of the
systems studied
in this paper is presented in Fig.~\ref{fig:fig1}. 
The metal atom
appears always displaced from the carbon layer. The height over the 
plane defined by its three nearest 
carbon neighbors is 
in the range 1.7-0.9~\AA. These three carbon atoms are also 
displaced
over the average position of the graphene layer by 0.3-0.5~\AA.
The total height (h$_z$) of the metal atom over the graphene plane
is the sum of these two contributions and 
ranges between 1.2-1.8~\AA, as shown in panel (c) of
Fig.~\ref{fig:fig2}. 
In most cases the metal atom occupies an almost perfect symmetric
configuration with  C$_{3v}$ symmetry. 
Exceptions are the noble metals,
that are slightly displaced from the central position, 
and Zn that suffers a Jahn-Teller
distortion in its most stable configuration. 
However, we have found that 
it is also possible
to stabilize a symmetric configuration for Zn with a
binding energy only $\sim$150~meV smaller. This
configuration was overlooked in a recently
published study on these systems~\cite{Krasheninnikov09}
and we will refer to it as Zn$_{\text{C}_{3v}}$ throughout
the paper.

Figures~\ref{fig:fig2} presents a
summary of the  
structural parameters 
of
substitutional 3$d$ transition metals, noble metals and Zn 
in graphene. Solid circles correspond to calculations
using the {\sc Siesta} code with pseudopotentials and a 
basis set of atomic orbitals, while open squares stand for
{\sc Vasp} calculations using plane-waves and PAW potentials.
The agreement between both sets of calculations is quite
remarkable. Data in these figures correspond
to calculations using a 4$\times$4 supercell of graphene.
For several metals we have also performed calculations
using a larger 8$\times$8 supercell and find almost identical
results. This 
is particularly true 
for the 
total
spin moments, which are 
less dependent
on the size of the supercell, but require a sufficiently dense 
k-point sampling to converge. 
In the following we will mainly discuss the 
results obtained with the smaller cell since the plots of the  
band structures are easier to interpret in 
that case.
Finally, as already mentioned, noble metals and Zn present
a distorted configuration. A detailed description of
the structural parameters in these cases will be given 
below, here we only present the averaged structural
data for noble metals and those 
corresponding to the Zn$_{\text{C}_{3v}}$ case.

The data in Fig.~\ref{fig:fig2} are basically consistent with
those reported by Krasheninnikov {\it et al.} in
Ref.~\onlinecite{Krasheninnikov09}.
The behavior of the metal-carbon bond length and
the height (h$_z$) of the impurity over the layer 
reflects approximately the size of the metal atom. 
For transition metals 
these distances decrease as we increase the atomic number,
with a small discontinuity when going from Mn to Fe.
The carbon-metal bond length reaches its minimum for 
Fe (d$_{\text{C-Fe}}$=1.76~\AA), keeping a very similar 
value for Co and Ni. 
For Cu 
and Zn the distances increase reflecting the fully occupied
3$d$ shell and the large size of the 4$s$ orbitals. Among the
noble metals we find that, as expected, 
the bond length largely increases
for Ag with respect to Cu, but slightly descreases
when going from Ag to Au. The latter behavior can be understood from
the compression of the 6$s$ shell due to 
scalar
relativistic effects.

\subsection{Binding energies}

The binding energies of the studied
substitutional metal atoms in graphene can be found in panel (d)
of Fig.~\ref{fig:fig2}. In general, the behavior of the 
binding energies can be correlated
with that of the carbon-metal 
bond length, although 
the former is somewhat more complicated. 
Binding energies for transition metals
are in the range of 8-6~eV. Ti presents the maximum binding energy,
which can be easily understood since for this element 
all the metal-carbon bonding 
states (see Sec.~\ref{sec:ScTi}) 
become fully occupied. One could expect 
a continuous decrease of the binding energy as we
move away from Ti along the transition metal series and first the
non-bonding 3$d$ and later the metal-carbon antibonding levels
become populated.
However, the behavior is non-monotonic and the smaller binding
energies among the 3$d$ transition metals 
are found for Cr and Mn, while a local maximum 
is observed for Co. This complex behavior is related
to the simultaneous energy down-shift 
and compression of the 3$d$ shell of the metal
as we increase the atomic number. This will 
become more transparent when discussing in detail
the metal-carbon hybridization levels.
In brief, the behavior of the binding energies
of the substitutional 3$d$ transition metal
comes from two competing effects: 

({\it i}) as 
the
3$d$ shell 
becomes occupied and 
moves to lower
energies the hybridization with the 
carbon vacancy states
near the Fermi energy (E$_F$) is reduced, which decreases
the binding energy; 

({\it ii})
the transition from Mn to 
late transition metals
is accompanied by a 
shift
of the metal-carbon bond length
of $\sim$0.1~\AA, 
which increases the carbon-metal interaction and, correspondingly,
the binding energy.

Binding energies for noble metals are considerably smaller
than for transition metals
and 
mirror the reverse behavior of 
the bond lengths:
3.69, 1.76 and 2.07~eV, respectively, for Cu, Ag and Au.
The smallest binding energy ($\sim$1~eV) 
among the metals studied here is found for 
Zn, withh filled $s$-$d$ electronic shells.

\subsection{Spin moments}
\label{subsec:moments}

The spin 
moments 
of
substitutional transition and noble metals in graphene
are
shown in Fig.~\ref{fig:fig3}.
Again they are in agreement with the results 
of
Ref.~\onlinecite{Krasheninnikov09}. However, 
we advance a 
simple model to understand the observed behavior 
which was not
presented in that reference. One of the fundamental results
of our study is a detailed model
of the bonding and electronic structure of
substitutional transition metals 
in graphene. As we will see below, 
the evolution of the spin moment
can be completely understood using such model. 
In brief, we can 
distinguish different regimes according 
to the filling of electronic levels of different
(bonding, non-bonding and antibonding) character: 

({\it i}) all the 
carbon-metal bonding levels are filled for
Sc and Ti and, correspondingly, the spin-moment is zero;

({\it ii}) non-bonding 3$d$ levels become
populated for V and Cr  giving rise to 
a spin moment of, respectively, 1 and 2~$\mu_B$ 
with a strong localized $d$ character;

({\it iii}) for Mn one additional electron is added 
to the antibonding $d_{z^2}$ level 
and the spin moment increases to 3~$\mu_B$;

({\it iv}) finally, for Fe and heavier atoms
all the non-bonding 
3$d$ levels are occupied and 
the spin moment oscillates between 
0 and 1~$\mu_B$ as
the
antibonding
 metal-carbon 
levels become occupied.

The sudden decrease of the spin moment from 
3~$\mu_B$ for Mn to 0~$\mu_B$ for Fe is
characterized by a transition from a complete 
spin-polarization of the non-bonding 3$d$ levels to 
a full ocuppation of those bands.
However, this effect
depends on the ratio between 
the effective electron-electron interaction 
within the 
3$d$ shell 
and the metal-carbon interaction (see Sec.~\ref{sec:FeMnborder}).
As we will see below, if the hybridization with the neighboring atoms
is artificially reduced, 
for example by increasing 
the Fe-C distance,
Fe impurities develop a spin moment
of 2~$\mu_B$. Our results also show that 
it is also possible to switch on the 
spin moment of Fe by changing the effective
electron-electron interaction within the 3$d$ shell.
This can be done using the
so-called GGA+U method. For a large
enough value of U (in the range 2-3~eV), Fe impurities
develop a spin moment of 1~$\mu_B$. 
This will be 
also
explained in detail in Sec.~\ref{sec:FeMnborder}.
For the time being we just 
point out that this behavior is unique to Fe:
using similar
values of U for other impurities does not
modify their spin moments. 

At the level of the GGA calculations
Fe constitutes the border between two different
characters of the spin moment associated with 
the substitutional metal impurities in graphene: 
3$d$ 
magnetism for V-Mn and 
a ``defective-carbon'' -like magnetism for heavier atoms.
For Co, Ni, the noble metals and Zn 
the electronic levels close to the
E$_F$ have a stronger contribution from 
the carbon nearest-neighbors and resemble the
levels of the isolated D$_{3h}$ carbon vacancy.
In particular, Mulliken population analysis show
that the spin moment of the noble metals impurities
has a dominant contribution for the three nearest carbon 
neighbors (see Table~\ref{tab:spinmoment}). 
For Zn two electrons occupy a two-fold degenerate
level reminiscent of the E $sp$ level of the unreconstructed
carbon vacancy (see Sec.~\ref{sec:D3hvac}). 
As a consequence, the system suffers a
Jahn-Teller distortion and has a zero spin moment.
However, it is possible to stabilize a symmetric
configuration (Zn$_{{\text C}_{3v}}$) 
with a moment of 2~$\mu_B$ and only
slightly higher in energy.

\section{Unreconstructed D$_{3h}$ carbon vacancy}
\label{sec:D3hvac}

We have seen in the previous summary of results that, 
as substitutional
impurities in graphene,
most of the 
metal atoms studied here present
a threefold symmetrical 
configuration.
For this reason we have found particularly instructive
to analyze 
their electronic structure as the result of the hybridization between the
atomic levels of the metal atoms
with the electronic levels associated with 
an unrelaxed
D$_{3h}$ symmetrical carbon vacancy.

Figure.~\ref{fig:fig4}~(a) shows the electronic
structure of such D$_{3h}$ carbon vacancy as calculated
using 4$\times$4 graphene supercell, while panel (b) 
presents a simplified scheme that highlights the 
defect levels associated with the vacancy 
and indicates their different character and symmetry.
The defect levels of the D$_{3h}$ vacancy can be
easily classified according to their $sp$ or $p_z$ character 
and whether they transform according to A or E-type 
representations. 
Close to the E$_F$ we can find a 
fully symmetric A $p_z$ level (thus belonging
to the A$^{\prime\prime}_2$ irreducible representation) 
and two degenerate (at $\Gamma$) 
defect levels with E symmetry and $sp$ character 
(E$^\prime$ representation). 
Approximately 4~eV below E$_F$ we find another defect level 
with A $sp$ character (A$^\prime_1$ representation).

It is interesting 
to note that it is not possible to identify any localized
defect level with E $p_z$ (E$^{\prime\prime}$) character. 
This is due to the strong coupling with the delocalized
states in the graphene layer and 
contrasts to the case of the A $p_z$ level.
The A $p_z$ level lies very 
close to E$_F$, where the density of states is low, and 
due to its A symmetry cannot appreciably couple
to the delocalized $p_z$ states of graphene 
in that energy range. On the contrary, the
E $p_z$ combinations present a very strong hybridization  
with the rest of the states of the graphene layer. 
Indeed, an inspection of 
the Projected Density of States (PDOS) 
(see the scheme in Fig.~\ref{fig:fig4}~(b)) 
reveals that the spectral weight associated with such E symmetry 
linear
combinations of p$_z$ orbitals of the carbon atoms
surrounding the vacancy 
extend over the whole valence and conduction band of graphene.
It is important to take 
the last observation
into account when developing a model
of the electronic structure 
for the metal substitutionals in 
graphene. 
Note that we need 
to have the correct number of electrons from 
carbon available for forming localized (covalent) bonds.

The three carbon atoms around the vacancy provide three unpaired
electrons associated with the unsaturated $sp$ lobes and 
three electrons coming from the $p_z$ orbitals.
As shown in Fig.~\ref{fig:fig4}~(b), 
two of these electrons stay in $p_z$ states delocalized
over the graphene layer while the other 
four electrons fill the A $sp$ and A $p_z$ levels 
localized at the vacancy.

\section{Analysis of the electronic structure}
\label{sec:discussion}

We now turn to the problem of the electronic structure 
of 3$d$ transition and noble 
metal atoms as substitutional impurities in graphene.
We first present a model of the hybridization between
carbon and metal levels and, subsequently, we show that this 
model allows to understand
in detail the band structures obtained in our calculation
for all the metals.

\subsection{Sc and Ti: filling the vacancy-metal bonding levels}
\label{sec:ScTi}

\begin{figure}
\includegraphics[width=3.250in]{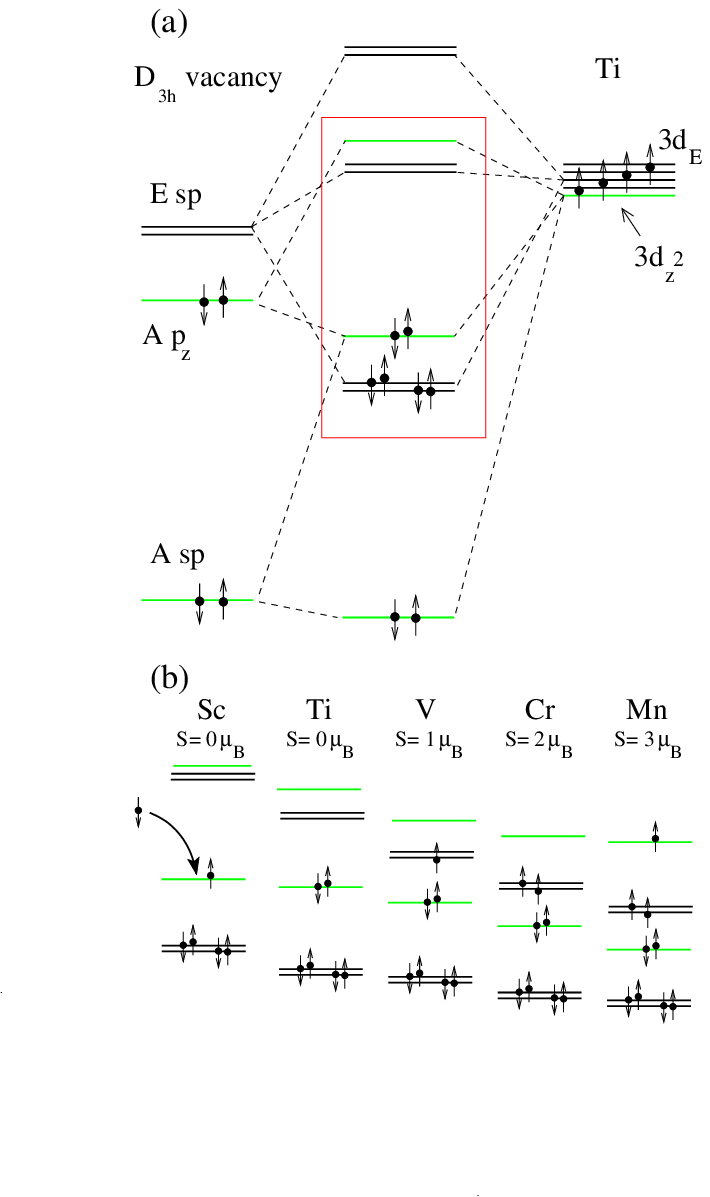}
\caption{\label{fig:fig5}(Color online) (a) 
Scheme
of the hybridization between the 3$d$ levels of Ti and 
the localized impurity levels of the D$_{3h}$ C vacancy. 
Only $d$ levels of Ti are taken into account since
our calculations show that, at least for transition metals,
the main contribution from $s$ levels appears well above E$_F$. 
Levels with A symmetry are represented by gray (green) lines,
while those with E symmetry are marked with black lines. The 
region close to E$_F$ is highlighted by a (red) square. (b)
Schematic representation 
of the evolution of the electronic
structure near E$_F$ for several substitutional transition
metals in graphene. The spin moment (S) is also indicated.
Substitutional Sc impurities act
as electron acceptors, causing the p-doping of the graphene layer.
}
\end{figure}

\begin{figure}
\includegraphics[width=3.250in]{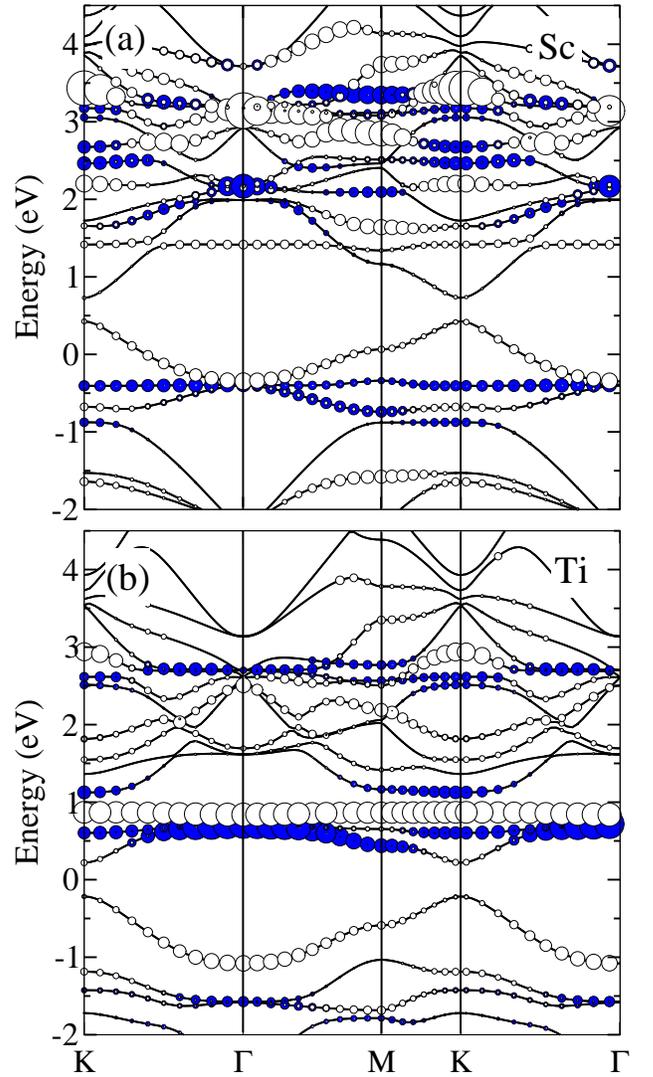}
\caption{\label{fig:fig6}(Color online) Calculated band structure
of substitutional Sc (a) and Ti (b) impurities in
4$\times$4 supercell of graphene. Open
circles indicate the contribution
from 3$d_{z^2}$ orbitals of the metal atom and
C 2$p_z$ orbitals of the neighboring C atoms. Solid circles
indicate contributions from the rest of the 3$d$ orbitals
and C 2$p_x$ and 2$p_y$ orbitals. Energies are referred to
the Fermi energy. }
\end{figure}

Figure~\ref{fig:fig5}~(a) presents a 
schematic representation  of the 
hybridization of the 3$d$ levels of Ti with those of 
an unreconstructed D$_{3h}$ carbon vacancy. 
We only consider explicitly the 
3$d$ states of the metal atom since our 
calculations 
show that,
for transition metals, the main contribution from 4$s$ orbitals
appears well above E$_F$. 
Due to the symmetric position of the metal atom over the vacancy 
the system has a C$_{3v}$ symmetry and the electronic 
levels can still
be classified according to the A 
or E irreducible representations of this point group.
Of course,
metal and carbon vacancy states
only couple when they belong to the
same irreducible represention. Thus, occupied
A $p_z$ and A $sp$ vacancy levels can 
only 
hybridize with
the 3$d_{z^2}$ orbitals (A$_1$ representation), while all the other 
3$d$ metal orbitals can 
only 
couple to the unoccupied E $sp$ vacancy 
levels.

With these simple 
rules in mind and taking into account
the relative energy position of carbon and metal levels, that
changes as we move along the transition metal series,
we can propose
a model of the electronic
structure of substitutional transition metals in graphene
as represented in Fig.~\ref{fig:fig5}~(a) and (b).
There are three localized defect levels with A$_1$ character
and three twofold-degenerate levels with E character.
Two of these E levels correspond to 
bonding-antibonding $sp$-$d$ pairs, while the third
one corresponds to  3$d$ non-bonding states.
For Sc-Mn the three A$_1$ levels can be pictured as
a low lying bonding level with A $sp$-$d_{z^2}$
character and a bonding-antibonding pair with 
A $p_z$-$d_{z^2}$ character. 

Therefore, as shown in Fig.~\ref{fig:fig5} we have
four metal-vacancy bonding levels 
(two A and one 
doubly-degenerate
E levels) that
can host up to eight electrons. 
Ti contributes
with four valence electrons, and there are 
four electrons associated with the 
localized carbon-vacancy levels.
Thus, for Ti the bonding 
states are completely occupied.
Consequently,
Ti presents the highest binding energy among all
3$d$ transition metals and has a zero spin moment.

The situation for Sc ougth to be discussed in detail. As Sc 
has three valence electrons, in principle
we could expect 
an incomplete filling of the metal-vacancy
bonding levels and a spin moment of $\sim$1~$\mu_B$.
However, in our model 
the highest bonding state (with A $p_z$-$d_{z^2}$ character)
appears below E$_F$ and the Sc impurity 
can act as an acceptor impurity.
Our
calculations show that this is indeed the case.
The Sc-vacancy  $p_z$-$d_{z^2}$ impurity level
captures an electron 
from the extended states of the graphene layer
which becomes p-doped. 
In total, the substituional Sc-graphene system
does not
show any spin polarization.

 We can now contrast the expectations from our model with 
actual calculations. Figure~\ref{fig:fig6} shows the
band structure of Sc (a) and Ti (b) close to the E$_F$.
As expected, the main contribution from the 3$d$ shell is 
found above E$_F$. Below E$_F$ we find one defect band with 
$p_z$-$d_{z^2}$ character and two bands (degenerate at $\Gamma$)
with $sp$-$d$ character. These bands are 
in close correspondence with the bonding A and E levels appearing 
in our model. 
In the case of Ti the E$_F$ is located inside a 
gap of $\sim$0.5~eV 
that opens at K point in the Brillouin zone. This gap appears
due to the relatively small 4$\times$4 supercell used
in these calculations and is reduced
when larger supercells are used. Thus, the filling 
of the graphene extended bands is not appreciably changed
by substitutional doping with Ti.
For Sc the situation is different. As shown 
in Fig.~\ref{fig:fig6}~(a) E$_F$
moves away from the K point, the Sc-vacancy complex
captures one electron and the graphene layer
becomes doped with holes. 

Regarding the unocuppied bands, 
the 3$d$ contribution for Sc
above E$_F$ appears quite broadened
due to the strong hybridization with the graphene
states. Indeed,
the defect levels are somewhat
difficult to identify and to correlate with our model.
One exception is 
a flat band with strong $d_{z^2}$ 
character appearing at $\sim$1.5~eV 
that, 
due to its symmetry,
does not couple so efficiently with the host states. 
The case of Ti is much
easier to interpret in terms of 
the simplified model presented
in Fig.~\ref{fig:fig5}~(a). In particular, we can find two 
bands at $\sim$0.6~eV
with strong $d_{xy}$ and $d_{x^2-y^2}$ contribution that correspond
with the non-bonding $d$ impurity levels, and one band 
with $d_{z^2}$ character at $\sim$0.8~eV corresponding 
with the A $p_z$-$d_{z^2}$ antibonding level.
Around 2.6~eV we can also find
the two E $sp$-$d$ antibonding defect 
levels, although in this case much more hybridized with the host.

\subsection{V, Cr and Mn: 3d magnetism}

\label{sec:VCrMn}

\begin{table}
\caption{\label{tab:spinmoment}
Mulliken population analysis of the spin moment 
in the central 
metal impurity (S$_M$) and the carbon nearest neighbors
(S$_C$) for different substitutional impurities in graphene. 
S$_{tot}$ is the total spin moment in the supercell.}
\begin{tabular}{|l|c|c|c|}
\hline
     &  S$_M$($\mu_B$)  & S$_C$ ($\mu_B$)  & S$_{tot}$ ($\mu_B$)  \\
\hline
V    & 1.21   & -0.09 & 1.0 \\ 
Cr     & 2.53 & -0.20 & 2.0 \\ 
Mn     & 2.91 & -0.10 & 3.0 \\
Co     & 0.44 & 0.06  & 1.0 \\
\hline
Cu     &
      0.24           
 & -0.03, 0.31, 0.31    
 &     1.0             \\
Ag   &      0.06             
     & -0.31, 0.54, 0.54          
     &  1.0            \\ 
Au  &    0.16          
    &   -0.28, 0.50, 0.50       
    &  1.0           \\
\hline
    Zn$_{\text{C}_{3v}}$  &    0.23
    &  0.37
    &  2.0    \\ \hline    
\end{tabular}

\end{table}

\begin{figure}
\includegraphics[width=3.250in]{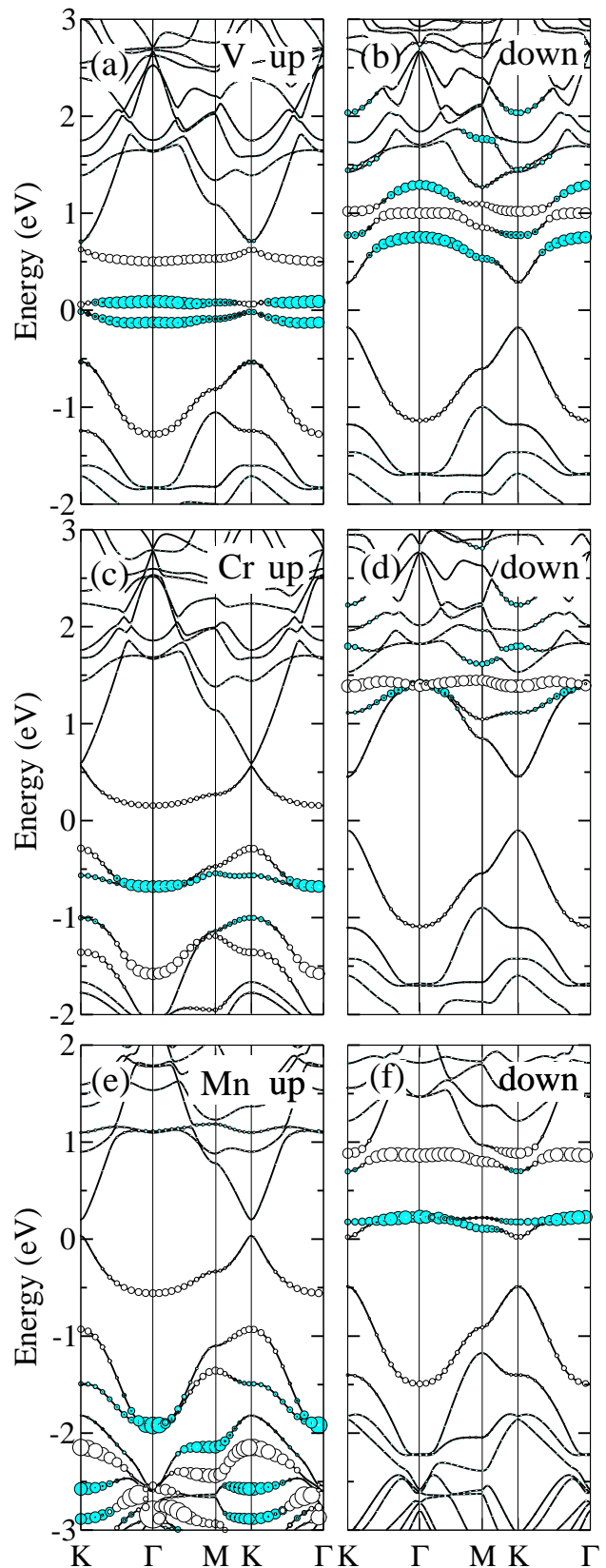}
\caption{\label{fig:fig7} (Color online)
Calculated band structure
of substitutional V (panels (a) and (b)), Cr ((c) and (d))
and Mn ((e) and (f)) impurities in a
4$\times$4 supercell of graphene. Open and filled
circles indicate, respectively, the contribution
from 3$d_{z^2}$ and the rest of the 3$d$ orbitals of the
metal and, therefore, also indicates levels
with A and E symmetries. Energies are referred to
the Fermi energy }
\end{figure}

As we have seen above, 
the metal-vacancy bonding levels are completely
filled for substitutional Sc and Ti and, as a consequence, 
the spin moment associated with these impurities is zero.
However, following our model in Figures~\ref{fig:fig5}~(a) and (b),
as we move along the transition metal 
series the
number of valence electrons increases 
and
the 
non-bonding 3$d$ impurity states start to become occupied.
Hence, due
to the strong atomic character and localization 
of these states,
the system develops a non-zero spin moment.

Figure~\ref{fig:fig7} shows the calculated band
structures for V, Cr and Mn impurities 
in a 4$\times$4 supercell
of graphene. 
The full calculations and the predictions of
our simplified model agree remarkably, 
at least in the neighborhood of
E$_F$.
For V and Cr 
this correspondence
is particularly 
evident:
one and two electrons, respectively, occupy the 
degenerate non-bonding E $d$ levels. These non-bonding
levels have a dominant contribution from the 3$d_{xy}$ and
3$d_{x^2-y^2}$ orbitals of the metal atoms.
As expected, the spin moments associated with these
impurities are, respectively, 1 and 2~$\mu_B$.
The strongly localized character of these spin moments is
corroborated by the Mulliken population analysis shown
in Table~\ref{tab:spinmoment}. This analysis indicates
that the spin moment is mainly localized at the 
metal impurity. 
The contribution from the neighboring carbon atoms is
much smaller and has the opposite sign.
The localized
character of the moment is also consistent with the relatively
large values of the spin splitting of the impurity bands.
From Figure~\ref{fig:fig7} we 
calculate a spin splitting 
of  $\sim$0.9~eV for V and almost 2~eV for Cr for the 
E $d$ levels at E$_F$.
These splittings are comparable to those of $d$-electrons 
in magnetic bulks, in the order of 1 eV.

Figure~\ref{fig:fig7}~(e) and (f) show the majority and
minority spin band-structures for Mn substitutionals
in graphene. 
In
addition to the  
non-bonding 3$d$ levels,  
the antibonding A $p_z$-$d_{z^2}$ defect level 
also becomes occupied and spin-polarized. This level
has an important  contribution from the 3$d_{z^2}$ state
of 
Mn (given by open symbols).  Therefore, it is relatively localized
within the Mn atom and
presents a significant tendency towards spin polarization.
Indeed, our calculations show that Mn substitutions in graphene 
give rise to a spin moment of
3~$\mu_B$. 
The Mulliken decomposition for Mn in Table ~\ref{tab:spinmoment}
confirms again the
localized character of this 
spin moment. However,
some differences respect to V and Cr are also
found. 
In those cases, the moment associated with the 
metal atoms was always  
larger that the total moment and the
only significant additional contributions came 
from the nearest carbon neighbors. 
However, for 
Mn the atomic
moment is somewhat smaller (2.91) than the total moment (3.0).
Taking into account the contribution from the 
nearest carbon neighbors (-0.3), a moment
of $\sim$0.4~$\mu_B$ is assigned to carbon atoms that are
further away from the 
defect. This indicates 
a slightly more delocalized
character 
 of the spin moment of Mn, since 
for V and Cr 
the ``long range'' contribution 
was  smaller than 0.1~$\mu_B$.
From the band structures in Fig.~\ref{fig:fig7} (e) and (f) we 
obtain a spin splitting of $\sim$2.1~eV for the non-bonding 
$d$ levels of the Mn impurity , similar to the case of Cr.
The spin splitting for the 
antibonding A $p_z$-$d_{z^2}$ state has a smaller value of $\sim$1.5~eV,
indicative of its larger spatial extension.

From the simple scheme presented in Fig.~\ref{fig:fig5}~(a)
we cannot completely determine the value of the 
spin moment of the Mn impurity.
It can be 3~$\mu_B$ as 
found in our first principles calculations
and schematically depicted in Fig.~\ref{fig:fig5}~(b). However, 
a magnetic moment of 1~$\mu_B$ is also a possible
answer. 
In the latter case,
the additional electron in Mn with
respect to Cr could populate one of the minority-spin
non-bonding $d$ impurity levels instead of the antibonding  
A $p_z$-$d_{z^2}$ level.
In such situation, the
spin moment is 
determined by a delicate balance 
between the 
onsite exchange energy  
within the 3$d$ shell, 
and 
the energy cost
($\Delta\epsilon_{Ad}$) 
to promote
one electron from the non-bonding $d$ levels
to the higher energy antibonding 
A $p_z$-$d_{z^2}$ state.
Note that the electron-electron repulsion is also reduced
when the electron moves into the less localized  A $p_z$-$d_{z^2}$ level.
An estimate of the exchange energy 
can be obtained from the spin splitting ($\Delta_S$)
of the defect levels nearby E$_F$.
The relative position of the 3$d$ states respect to 
the A $p_z$ level of the carbon vacancy 
and the interaction matrix element
between these levels determine $\Delta\epsilon_{Ad}$.
Thus, within our GGA-DFT calculations we can expect 
the high spin solution to be
favored approximately when $\Delta_S$ $>$ $\Delta\epsilon_{Ad}$.
From the band structures 
in Fig.~\ref{fig:fig7}~(e) and (f)  we obtain  
$\Delta\epsilon_{Ad}\sim$~1.0~eV, which is 
smaller than the values of $\Delta_S$ discussed
previously and, therefore, is consistent with 
the calculated moment of
3~$\mu_B$ for Mn.
Fe impurities considered 
below in detail present 
a similar
situation where two spin configurations are possible. However,
in the case of Fe the low spin (spin compensated)
solution is preferred at the level of DFT-GGA calculations
as a result of the stronger metal-carbon hybridization.

In short, both the results of the calculations and
the expectations based on our model of 
the metal-vacancy bonding point towards a very strong 3$d$
character  of 
the defect levels  appearing nearby
of E$_F$ for V, Cr and Mn substitutional impurities 
in graphene. 
The filling of 
these localized
levels favors high spin solutions in accordance 
with the first Hund's rule of atomic physics.
Thus,
we can picture the appeareance of spin polarization for 
V, Cr and Mn substitutionals in graphene as 
``standard'' $d$-shell magnetism.

\subsection{Fe, Co, Ni: strong contribution from the carbon vacancy levels}
\label{sec:FeCoNi}

\begin{figure}
\includegraphics[width=3.250in]{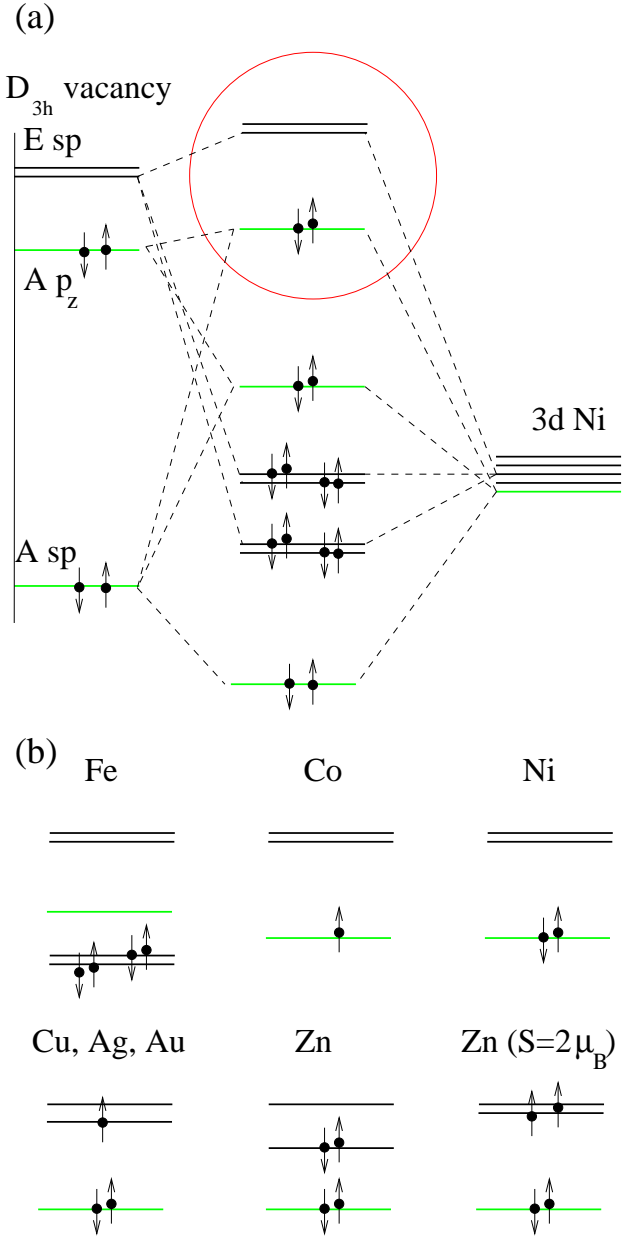}
\caption{\label{fig:fig8}
(Color online) (a) Similar to Fig.~\ref{fig:fig5} (a)
for the case of Ni. The region close to the Fermi level 
is indicated by a (red) circle. 
(b)
Scheme
of the 
levels close the E$_F$ for Fe, Co, Ni, noble metals and Zn.
For Fe, in addition to the antibonding metal-vacancy levels,
we have also included the non-bonding $d$ levels that also 
appear quite close to E$_F$.
The noble metals slightly break the C$_{3v}$ symmetry.
For Zn there are two solutions: a high spin solution that preserves
symmetry and a distorted one with zero spin moment.}
\end{figure}

\begin{figure*}
\includegraphics{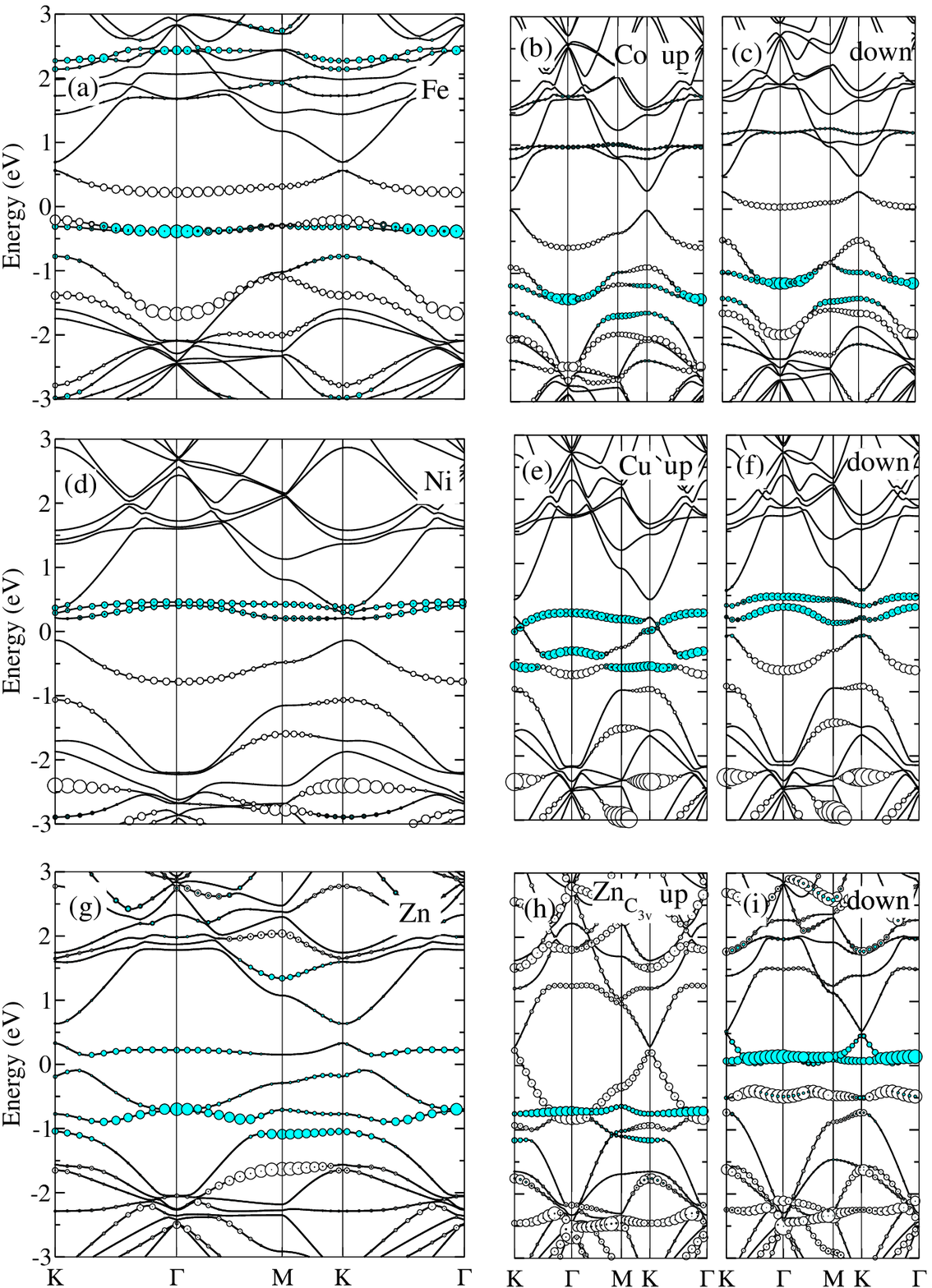}
\caption{\label{fig:fig9}(Color online) Like Fig.~\ref{fig:fig7} for 
Fe (a), Co (b) and (c), Ni (d), Cu (e) and (f), Zn (g) and symmetric Zn$_{\text{C}_{3v}}$ (h) and (i) substitutional
impurities in graphene.}
\end{figure*}

As we have seen in the previous section the defect levels 
appearing in the neighborhood of E$_F$ 
associated with the presence of
V, Cr and Mn substitutional impurities in graphene 
have a strong 3$d$ 
character. Consequently,
these impurities
exhibit large spin moments.
However, 
when increasing the
atomic number along the transition series, 
the atomic 3$d$ 
levels move
to lower energies 
and we enter a different 
regime: the
defect states nearby E$_F$ 
have a predominant contribution from the carbon atoms neighboring 
the metal impurity. 
For 3$d$ 
late
transition metals heavier than Fe, the noble metals
and Zn we can stablish 
a strong link
between the electronic structure of these impurities around
E$_F$ and that of the unreconstructed D$_{3h}$ carbon vacancy.

A detailed scheme of the 
hybridization of the Ni 3$d$ states with those of
the carbon vacancy, and
the resulting electronic structure, is presented in 
Fig.~\ref{fig:fig8}~(a).
All the bonding and non-bonding metal-vacancy 
states are filled in this case and the levels
appearing  closer to E$_F$ have
antibonding character and a strong contribution
from the three carbon nearest-neightbors. 
Closely below 
E$_F$ we find a level with A character and a strong C 2$p_z$ 
contribution and a smaller weight in the Ni 3$d_{z^2}$ state.
Above $E_F$ there are two degenerate levels with E character
mainly coming from the C 2$sp$ lobes 
and hybridized with the Ni 3d$_{xz}$ and 3d$_{yz}$ orbitals.
The resulting electronic structure
strongly resembles that of the
isolated D$_{3h}$ (unreconstructed)
carbon vacancy 
as  can be checked by
comparing the band structures in Fig.~\ref{fig:fig4}~(a) and
Fig.~\ref{fig:fig9}~(d).
The main difference stems from the slightly higher position
of the unoccupied E levels in the case of the Ni impurity.
This upward shift is due to the antibonding interaction
with the $d$ states of Ni and contributes to the stability
of the C$_{3v}$ spin-compensated solution. In the
case of the D$_{3h}$
carbon vacancy, the E C 2$sp$ levels lie closer to 
E$_F$ and make the system unstable against spin and structural
distortions. 
This clear connection
between the electronic structure of Ni and that of the D$_{3h}$
vancancy was already emphasized in Ref.~\onlinecite{Santos08}.

Thus, our GGA calculations predict Ni substitutional to be 
a closed shell system with zero spin moment. 
One could expect that a better description 
of the 
electron-electron interaction within the 3$d$ shell 
would enhance any tendency of the system 
towards a magnetic instability.
For this reason we have performed GGA+U 
calculations (using 
the {\sc Vasp} code) 
with values of U up to 4.5~eV. 
However, Ni substitutional 
impurities in {\it flat} graphene always
remain non-magnetic in the calculations.

In Ref.~\onlinecite{Santos08} we proposed a different
way to switch on the magnetism of Ni substitutionals:
the possibility of spin-moments induced by curvature.
The idea 
consists in lifting the
degeneracy of the two unoccupied E $sp$-$d$ levels
by applying a 
structural distortion. If the distortion is large
enough 
one of these levels 
close to E$_F$ becomes partially populated
and, due to its small band width,  
spin polarized. We have
checked that the curvature of the carbon layer 
in (n,n) nanotubes with n ranging between 4 and 8 
induces spin moments
as large as 0.8~$\mu_B$ per Ni substitutional impurity. 
The spin moment in these substitutionally Ni-doped
nanotubes is strongly dependent, not only on the 
layer curvature, but on the 
density and arrangement of defects in the tube. 
We have recently demonstrated that 
a similar switching of the magnetic moment 
can be obtained in flat Ni-doped graphene 
by applying adequate
structural distortions,  thus providing a 
simple way to control the spin of  this system.~\cite{Portal09}

As suggested by our scheme in Fig.~\ref{fig:fig8}~(b),
we can now try to understand
the electronic structure of Co and Fe impurities
from that of the Ni substitutional but removing, repectively,
one and two electrons.  
According to this image
Co substitutionals should present
a spin moment of 1~$\mu_B$, while Fe substitutionals should be 
non-magnetic. 
This is indeed confirmed by our GGA calculations.

Figures~\ref{fig:fig9}~(b) and (c)
show, respectively, the majority and minority spin band
structures of the Co impurity. In the neighborhood of
E$_F$ we find a spin-polarized band associated
with the antibonding A $p_z$-$d_{z^2}$ impurity level.
The spin splitting of this band is $\sim$0.5~eV.
The hybridization character of this level
is confirmed by the Mulliken analysis
in Table.~\ref{tab:spinmoment}. 
Only a contribution of
0.44~$\mu_B$ to the total spin moment comes
from the Co atom. The relatively delocalized
character of the A $p_z$-$d_{z^2}$ level also becomes evident.
Only a moment of 0.18~$\mu_B$ comes from
the three carbon nearest neighbors, while 0.38~$\mu_B$ 
comes from carbon atoms at larger distances. 
The slow distance decay of the A $p_z$-$d_{z^2}$ 
defect level translates into quite strong and long-range 
magnetic interactions between moments
associated with neighboring Co defects.~\cite{Santos09} 
Indeed,
the peculiar
electronic structure of the Co impurities has important 
consequences for the magnetism of this system: 
couplings show a complex dependence with distance and direction, 
while the total spin-moment is determined by 
the number of Co substitutions in each 
sublattice of the graphene layer. We refer the interested
reader to Ref.~\onlinecite{Santos09}. 

Figure~\ref{fig:fig9}~(a) presents the GGA band structure for 
a Fe substitutional defect
in a 4$\times$4 supercell of graphene.
Similar results are found using a larger 8$\times$8 supercell.
This band structure is again in
reasonable agreement with the
simple model presented in Fig.~\ref{fig:fig8}. The non-bonding
$d$ levels are completely filled and appear 
$\sim$0.4~eV below
E$_F$ in the vicinity of $\Gamma$. The 
A $p_z$-$d_{z^2}$ level is mostly unoccupied and
close to  
0.2~eV above E$_F$ near $\Gamma$. However, we can see 
that nearby the K point the $p_z$-$d_{z^2}$ band 
becomes partially occupied, indicating a small charge 
transfer from the dispersive
$\pi$ bands of graphene to the defect. Mulliken 
analysis also reflects a small charge accumulation 
of $\sim$0.16 electrons in Fe. In spite of this
small partial population, the spin compensated
solution is the most stable for Fe substitutionals at 
the GGA level.

As already pointed out in Sec.~\ref{subsec:moments},  
the magnetic behavior
of the Fe imputiy is a consequence of a delicate balance
between the onsite electron-electron interaction and 
the metal-carbon hybridization. For this reason, we devote
a whole Section below
(Sec.~\ref{sec:FeMnborder}) to explore
how the band structure and spin moment of Fe substitutionals
are modified when these factors are independently controlled
by changing the Fe-C bond length and using the GGA+U approximation 
to describe the effects of the electron-electron repulsion 
within the $d$-shell. 

\subsection{Noble metals}
\label{sec:noble}

\begin{table}
\caption{\label{tab:noblemetals} 
Structural parameters for substitutional 
noble metals in graphene.
}
\begin{tabular}{|l|c|c|c|}
\hline
& d$_{\text{C-M}}$(\AA) & h$_z$(\AA) 
& $\theta$ ($^{\circ}$)  \\ \hline
Cu &  1.93, 1.90, 1.90     
  & 1.40                          
   & 88.9, 88.9, 95.2  \\

 Ag &  2.23, 2.19, 2.19      
     & 1.84                
      &  71.7, 71.7, 76.7    \\
 Au & 2.09, 2.12, 2.12   
  &   1.71       
      & 78.0, 78.0, 81.6  \\ \hline
\end{tabular}

\end{table}

\begin{figure}
\includegraphics[width=3.250in]{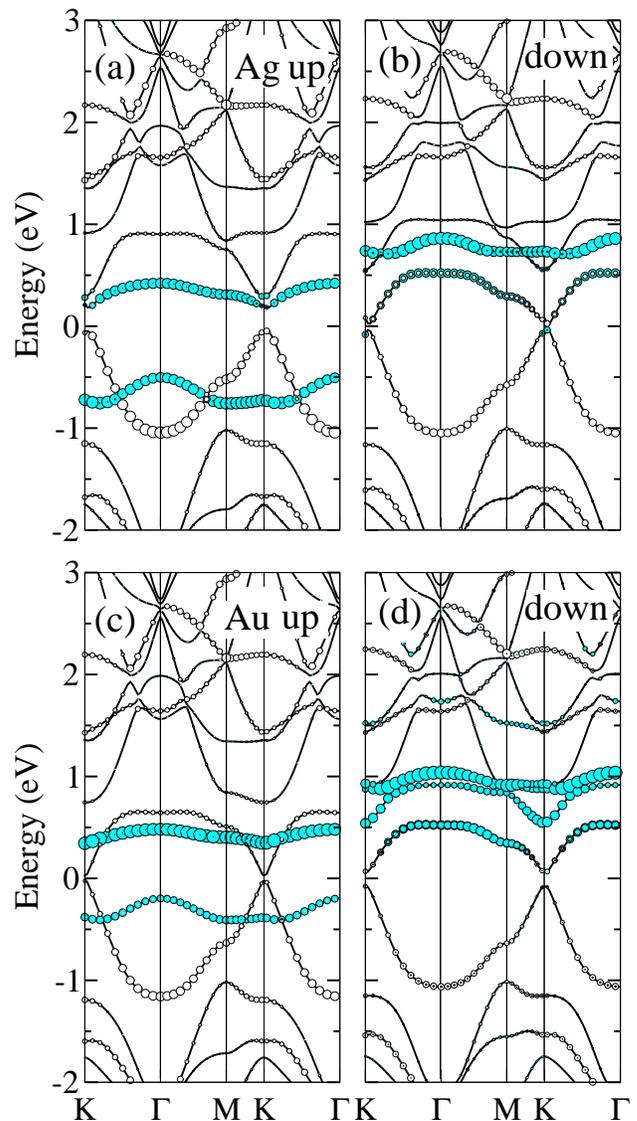}
\caption{\label{fig:fig10} (Color online)
 Like Fig.~\ref{fig:fig7} for
Ag and Au. The similarities with the band structure
of Cu in Fig.~\ref{fig:fig9} are evident.}
\end{figure}

In the previous sections we have seen
that some of the traditional ferromagnets, like Fe and
Ni, become non-magnetic as substitutional impurities
in graphene.  
Also quite surprisingly, we find here 
that substitutional impurities
of noble metals are
magnetic 
with a spin moment of 1~$\mu_B$. The reason 
for this behavior becomes clear once the electronic structure
of these defects is understood. 

The 
band structures can be found in 
Fig.~\ref{fig:fig9}~(e) and (f) for Cu and in Fig.~\ref{fig:fig10}
for Ag and Au. 
In agreement with the predictions of our simplified
model in Fig~\ref{fig:fig8}, we 
can see that in the case of the 
noble-metal impurities
the two-fold degenerate
E $sp$-$d$ antibonding levels~\cite{nomenclature} are now occupied 
with one electron. 
The system
undergoes a small structural distortion that removes the 
degeneracy of these levels and the unpaired electron becomes spin polarized. 
Therefore, substitutional impurities of the noble metals in graphene 
exhibit a spin moment of 1~$\mu_B$.
Relativistic effects are known to be much more important for Au
than for Ag and Cu. Although
we have not included spin-orbit coupling in our calculations,
scalar relativistic effects are taken into account in
the construction of the pseudopotentials. However, 
the similarities between the electronic structure
of all three noble metals are evident, which indicates
that bonding and magnetic behavior is mainly
dictated by the number of valence electrons. 

Structural parameters for all noble
metals impurities can be found in Table~\ref{tab:noblemetals}.
For Cu and Ag one of the metal-carbon bond lengths
is slightly larger than the other two,
whereas for Au one is shorter than the others. However, the
distortions are rather small with variations of
the bond lengths smaller than 2~\%.
The differences introduced by the larger scalar
relativistic effects of Au mainly reflect
in the slightly smaller metal-carbon bond length
for this metal as compared with Ag. 

Table~\ref{tab:spinmoment} shows
the distribution of the spin moment among
the metal atom and the nearest carbon neighbors. We
can see that the 
contribution from the
metal atom is almost negligible, particularly
in the case of Ag. This can be expected from 
the lower energy position of the $d$ shell in the
case of the noble metals as compared with transition metals. 
Although slightly 
hybridized with the $p$ shell of the metal
impurity, the defect states nearby E$_F$
in this case are mainly coming from the 
carbon 
neighbors.
Still the spin moment is rather localized in the complex formed
by the metal atom and its three nearest neighbors, which 
for Cu and Ag contributes
with a moment of 0.83~$\mu_B$, and up to 0.88~$\mu_B$ 
for Au. The
contribution from the rest of the graphene layer
is much smaller than, for example, in the case of Co. This reflects the 
dominant contribution from the relatively localized
carbon $sp$ lobes in the description of the defect states
nearby E$_F$ for these impurities.
This analysis reinforces the
link with the electronic structure
of the unreconstracted D$_{3h}$ carbon vacancy in graphene
as presented in Sec.~\ref{sec:FeCoNi}.
In fact, we can picture the main role of late transition- and 
noble-metals 
substitutionals in graphene as to stabilize the structure
of the carbon monovacancy, that otherwise will severely reconstruct,
and to change its charge state.

\section{Jahn-Teller distortion of substitutional Zn}
\label{sec:Zn}

\begin{table}
\caption{\label{tab:Zn}
Structural parameters and binding energies
for substitutional
Zn impurities in graphene for the symmetric C$_{3v}$
and most stable distorted configurations.
}
\begin{tabular}{|l|c|c|c|c|}
\hline
 & E$_B$ (eV)& d$_{\text{C-M}}$ (\AA) & h$_z$ (\AA) 
& $\theta$ ($^{\circ}$)  \\ \hline
Zn$_{\text{C}_{3v}}$ & 0.91 & 1.99
  & 1.67
   & 87.9  \\
Zn & 1.07 &  2.06, 1.89, 1,89
     & 1.54
      &  88.3 88.3 103.9    \\ \hline
\end{tabular}

\end{table}

For Zn impurities a second electron is added to the two-fold 
degenerate
E $sp$-$d$ shell. 
Under these circumstances two scenarios 
are possible: ({\it i}) a non-magnetic solution 
in which the system has undergone a 
Janh-Teller-like distortion, 
or ({\it ii}) a high-spin solution that maintains
the symmetric C$_{3v}$
geometry of the defect.
The relative energy of both solutions depends on the 
balance between the energy gain associated with the
distortion and the exchange energy of the electrons.
Both types of 
solutions are obtained in our density functional 
calculations for Zn substitutional impurities in 
graphene.

The details of the structure and the
binding energies of the Zn impurity are presented
in Table~\ref{tab:Zn} as calculated with {\sc Siesta}. 
Very similar results are obtained for both configurations
using {\sc Vasp}. The distorted configuration 
presents one larger Zn-C bond 
(by $\sim$3.5\%) and two shorter bonds ($\sim$5\%) 
compared with the bond length (1.99~\AA) of the undistorted
geometry.
The distorted configuration is more stable by 
160~meV (120~meV using {\sc VASP}).
This rather small energy difference between the two 
configurations might 
point to the appearance of 
non-adiabatic
electronic effects at room temperature.

The band structure for both configurations of Zn 
substitutionals can be found in Fig.~\ref{fig:fig9}.
Again they 
confirm the model presented
in Fig.~\ref{fig:fig8}. The distorted Zn 
[Fig.~\ref{fig:fig9}~(g)] breaks the degeneracy of
the E $sp$-$d$ levels: one of them appears fully occupied
$\sim$0.8~eV below E$_F$, while the other appears
a fews tenths of eV above E$_F$. For the C$_{3v}$ Zn impurity 
both E $sp$-$d$ bands are 
degenerate and the splitting 
between majority and minority levels
is $\sim$0.71~eV. Table~\ref{tab:spinmoment} shows the
Mulliken population analysis of the spin moment
for the  C$_{3v}$ Zn systems.
As in the
case of the noble metals the contribution from
the three nearest carbon neighbors is the most important.
However, the contribution of 
Zn is somewhat larger
and the total spin moment is more delocalized
with a contribution form the rest of the graphene layer
of $\sim$0.66~$\mu_B$. We should note that here
we have one additional electron as compared to the noble metal systems.

\section{Fe substitutionals: competition between
intra-atomic interaccions and metal-carbon hybridization}
\label{sec:FeMnborder}

\begin{figure}
\includegraphics{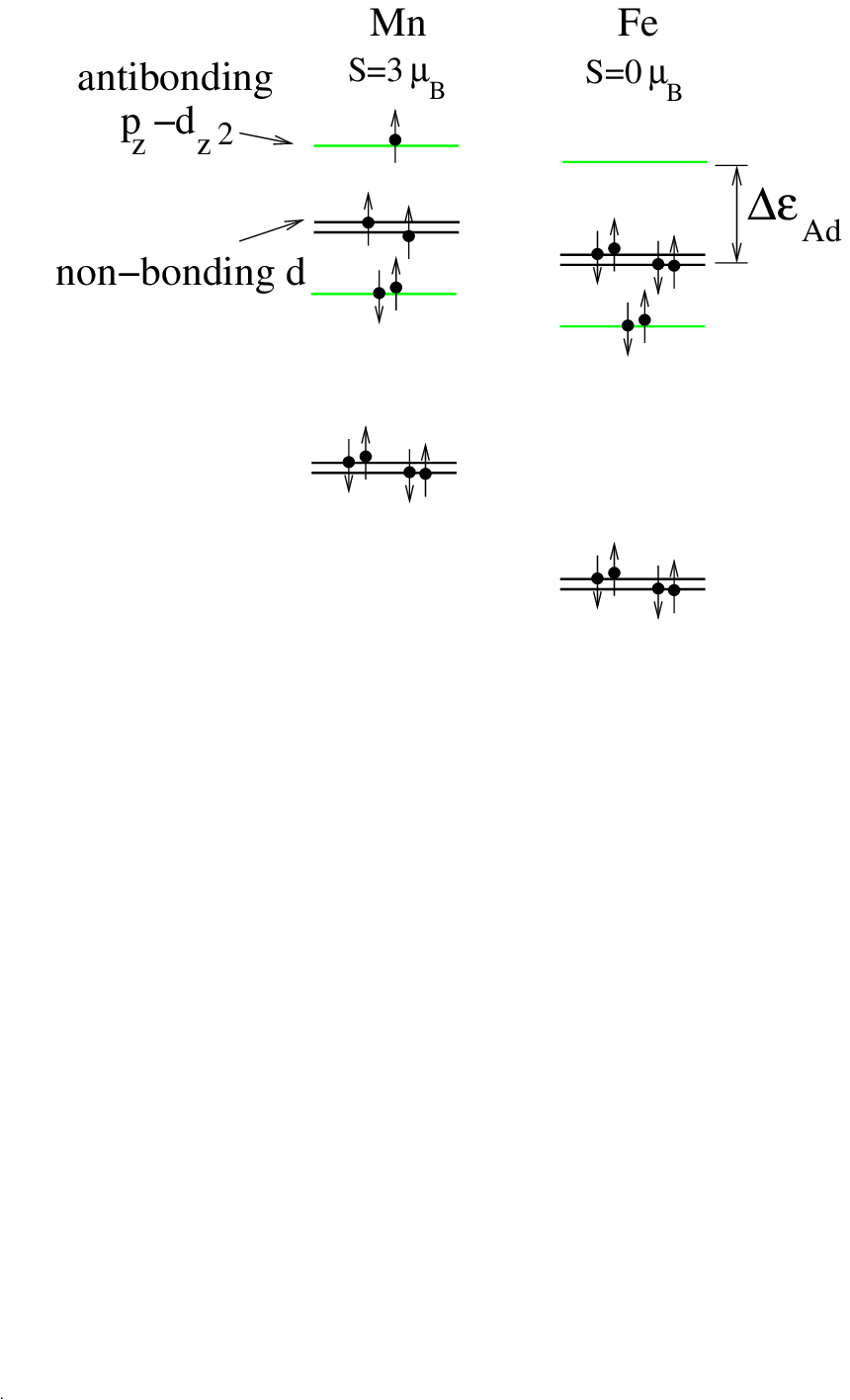}
\caption{\label{fig:fig11}(Color online)
Scheme of the electronic levels near E$_F$ for
Mn and Fe substitutionals in graphene as
deduced from our model of bonding and 
GGA calculations. 
$\Delta\epsilon_{Ad}$ is the 
energy cost to promote an electron to
the antibonding A $p_z$-$d_{z^2}$ 
hybridization level 
from the
non-bonding 3$d$ states. The magnitude of $\Delta\epsilon_{Ad}$,
relative to that of the 
spin-splitting 
$\Delta_S$
of these defect
levels, is crucial to determine the spin state 
of these impurities.
}
\end{figure}

\begin{figure}
\includegraphics{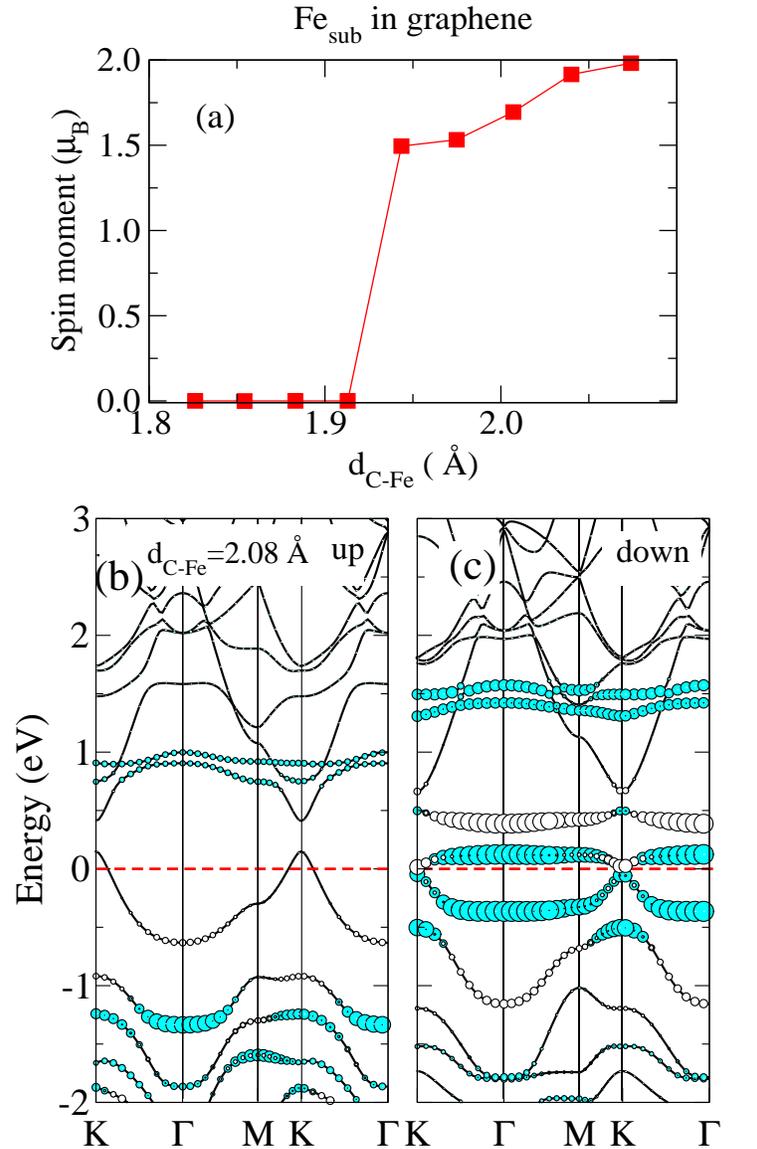}
\caption{\label{fig:fig12}(Color online) Panel (a) shows the 
spin moment of a substitutional Fe impurity in graphene
as a function of the C-Fe bond length.
Equilibrium position corresponds
to d$_{\text{C-Fe}}$=1.78~\AA. Panels (b) and (c) 
show the band structure for d$_{\text{C-Fe}}$=2.08~\AA.
Symbol code similar to Fig.~\ref{fig:fig7}. Energies referred to
the Fermi level (indicated by a dashed line). } 
\end{figure}

We have already pointed out that Fe substitutionals in graphene 
occupy a rather special
place at the border between two well 
defined regimes (see Fig.~\ref{fig:fig3}): 
({\it i}) the 
strong 3$d$ character of the defect levels nearby E$_F$
and large spin moments found for V, Cr and Mn
impurities, and ({\it ii}) the larger carbon character of those
electronic
levels and the small
oscillatory spin-moments of Co,  Ni and the noble metals.
GGA calculations  [Fig.~\ref{fig:fig9}~(a)] locate Fe impurities
within the second group,  with all the 3$d$ non-bonding 
levels fully occupied
and non-magnetic.
Thus, the spin moment
drops from 3~$\mu_B$ for Mn impurities to zero for
Fe, showing a quite discontinuous
behavior as a function of the number of
valence electrons.
In analogy with the standard Slater-Pauling
rule for transition metals, one could expect to find a more
gradual decrease of the spin moment as the number of valence
electrons is increased, i.e. Fe would 
have
a moment of 2~$\mu_B$.
In the present section we study in detail why 
the non-magnetic solution is more stable for Fe.

\subsection{ Key parameters: metal-carbon hopping and intra-atomic 
Coulomb interactions}

Figure~\ref{fig:fig11} shows a scheme of the electronic structure
of both Mn and Fe defects. Depending on 
how electrons are arranged among the A $p_z$-$d_{z^2}$ antibonding
level and the non-bonding 3$d$ 
states, Mn
can exhibit spin moments of 
3~$\mu_B$ or  1~$\mu_B$, while Fe
can have a moment of 
2~$\mu_B$ or can be  non-magnetic.
At the GGA level 
Mn prefers the
high-spin configuration, while the low-spin one is more
stable for Fe.
As commented in 
Section~\ref{sec:VCrMn}, the relative stability of the 
different spin states depends on the balance between
the effects of Coulomb repulsion 
and exchange within the 3$d$ levels
and the relative energy position of
the impurity levels given by $\Delta\epsilon_{Ad}$ 
(see Fig.~\ref{fig:fig11}).
The  hybridization with the neighboring 
C atoms is crucial in this interplay since it influences
({\it i}) the degree of localization of the defect levels
and the screening of 
Coulomb interactions, which modify the
spin splitting of the electronic levels $\Delta_S$, 
and ({\it ii}) 
the value of  $\Delta\epsilon_{Ad}$
through 
the effective hopping parameter between the 3$d$ states
and the A $p_z$ carbon vacancy level. 
Next, we shall deal with both aspects for subtitutional Fe in graphene.

\subsubsection{ Changing Fe-graphene hopping with distance}

By tuning the interaction 
with the host structure it should be possible 
to change the spin moment of these impurities.
We can modify the hopping by artificially changing 
Fe-graphene distance. 
The results of such calculation are shown in Fig.~\ref{fig:fig12}~(a).
While maintaining the
three-fold symmetry of the system, we 
have performed a series 
of calculations by progressively 
increasing the height of the Fe atom
over the graphene 
layer.
Increasing the C-Fe bond length by 
$\sim$9\% we observe 
an abrupt jump of the spin moment from zero to 1.5~$\mu_B$. 
The spin moment continues to rise and saturates 
at a value of  2.0~$\mu_B$
for d$_{\text{C-Fe}}$$\sim$2.07~\AA. This convincingly shows
that the metal-carbon hybridization is the key parameter
to explain the non-magnetic state of Fe substitutional impurity
in graphene.
When
increasing the C-Fe distance we 
mainly decrease $\Delta\epsilon_{Ad}$. Thus, we reduce the 
energy penalty for promoting electrons from the non-bonding $d$
to the A $p_z$-$d_{z^2}$ defect levels. 
At the same time we also increase the atomic character
of the non-bonding $d$ states and reduce the effect
of the screening due to the electrons in the graphene layer.
This $\Delta\epsilon_{Ad}$ reduction promotes
the electron-electron repulsion 
within the 3$d$ states of Fe and, eventually, stabilizes the 
solution with  2.0~$\mu_B$. 

\begin{figure}
\includegraphics{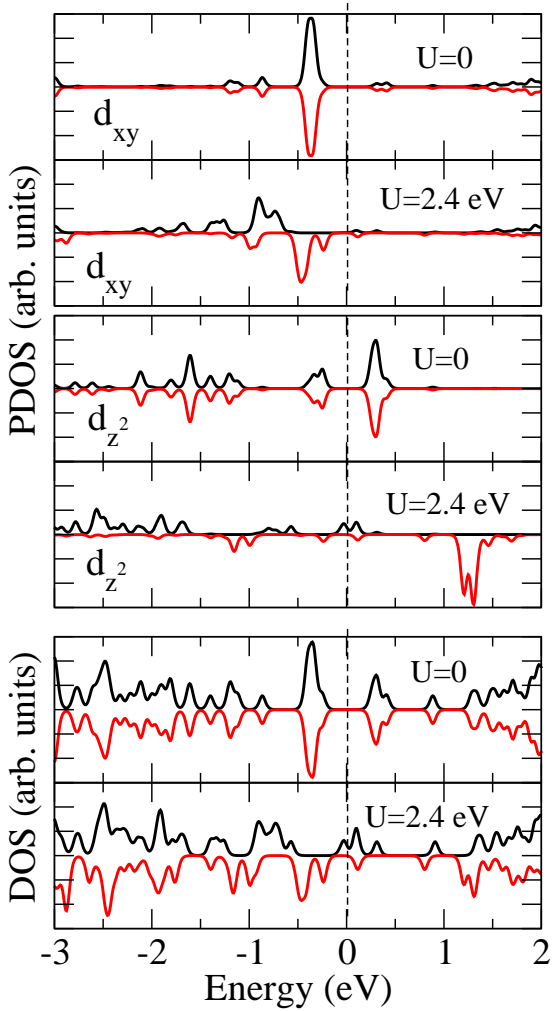}
\caption{\label{fig:fig13} (Color online)
Projected density of states (PDOS) and total density of 
states (DOS) for a Fe substitutional impurity in graphene
calculated  with {\sc VASP} 
using GGA and GGA+U with U=2.4~eV. Positive (negative) values
for majority (minority) spin. 
When U=0 the system shows zero spin-polarization and 
the Fe 3d$_{z^2}$ state is partially occupied. A large
enough Hubbard U forces an integer population of the 
Fe 3d$_{z^2}$ state and produces a spin moment of 1~$\mu_B$.}
\end{figure}

Figures~\ref{fig:fig12}~(b) and (c) present the band
structure for the high-spin state of
Fe, where the C-Fe bond length has been 
elongated up to 2.08~\AA.
We notice
the differences with the spin-compensated
ground state in Fig.~\ref{fig:fig9}~(a). On the one hand,
one electron
is promoted from the non-bonding $d$ states to the more delocalized
antibonding A $p_z$-$d_{z^2}$ impurity level in order
to reduce the effect of electron-electron repulsion. On the other
hand, the 
spin moment is maximized
in accordance with Hund's first rule. 

Therefore, we have seen that the
non-magnetic character of the Fe substitutionals
in the GGA calculations is due to the larger
interaction with the 
graphene layer,  
as compared for example with the Mn impurity. 
This
is consistent with the fact that Fe, together with 
Co and Ni, presents the smaller carbon metal bond length
among the whole series of 3$d$ transition metals. Fe impurities
also have one of the largest binding energies.
By artificially reducing this interaction
it could be
possible to obtain 
magnetic Fe substitutional impurities in graphene.

\subsubsection{Changing intra-atomic Coulomb interaction U}

Another route 
to explore would be
to increase the size of 
intra-atomic electron-electron interactions.
We have done that by using the so-called GGA+U methodology 
in which a Hubbard term is explicitly added to the DFT Hamiltonian
and solved within  the mean-field approximation. 
Our results indicate that with a reasonable
value of U ($\sim$2~eV or larger)
we obtain 
a magnetic solution for Fe substitutionals. However, 
contrary to our initial expectation this solution does not
correspond to the 2~$\mu_B$ high-spin solution discussed above, 
but to a new solution with 1~$\mu_B$. 
The key reason to understand this behavior is the partial 
occupation of the Fe 3$d_{z^2}$ state at the level of GGA calculations,
The $d_{z^2}$ state is strongly coupled to 
the delocalized A $p_z$ vacancy level appearing nearby E$_F$ and 
both, this hybridization and the population of the atomic orbital, 
are strongly modified
when the value of U is increased. 

Figure~\ref{fig:fig13} shows the results of the electronic
structure of Fe impurities calculated using
GGA+U with  U=2.4~eV and compared with those with U=0.
The two upper panels
show the projected density of states (PDOS) 
onto the $d_{xy}$ orbital (the projection onto
the $d_{x^2-y^2}$ orbital is identical by symmetry),
the two middle panels show the PDOS onto the d$_{z^2}$ orbital
of Fe and the two lower panels the total density of states.
At the level of GGA, 
with U=0, there is a very well
defined peak closely below E$_F$ in the 
$d_{xy}$ PDOS
corresponding to the position of the non-bonding $d$ states. 
The spectral weight coming from  the  $d_{z^2}$ orbital
is more spread: presents some broad structure around 
$-1.5$~eV corresponding to the bonding A $p_z$-$d_{z^2}$ defect level
interacting with the valence band of graphene, and a narrowers peak
near E$_F$ with origin in the slightly occupied 
antibonding A $p_z$-$d_{z^2}$ impurity state. Therefore,
from the Fe 3$d$ states appearing nearby E$_F$ we can picture
the $d_{xy}$ and $d_{x^2-y^2}$ states as fully occupied and
the $d_{z^2}$ as partially occupied due to the larger
interaction with the carbon neighbors.

For a sufficiently
large value of U the C 2$p_z$-3$d_{z^2}$ hybridization 
is overcome by the tendency of the 
electrons to have an integer population
within the
localized 3$d$ shell. Thus, the 3$d_{z^2}$ 
localizes one electron and,  as a consequence, 
the system develops a 1~$\mu_B$
spin-polarization.
This changes can be appreciated in 
Fig.~\ref{fig:fig13}.
The 
majority-spin 3$d_{z^2}$ PDOS shifts downwards,
while a strong  unoccupied peak
appeas around 1.5~eV in the 3$d_{z^2}$ minority spin PDOS. 
Simultaneously, a more delocalized
level mostly coming from the 2$p_z$ orbitals of
the neighboring C atoms appears half-filled at E$_F$.
Note that other levels, such as the $d_{xy}$, do not suffer
such strong modifications since they are already almost fully
occupied with U=0. It is also interesting to note that, 
due to symmetry considerations, they couple with 
the graphene layer very differently 
from the $d_{z^2}$ level. 

\subsection{Revelance for recent experiments
of Fe implantation in graphite}

There are recent reports on the paramagnetism of  
iron-implanted graphite
that indicate the existence of local magnetic moments
associated with the implanted Fe atoms.~\cite{Sielemann08,Barzola08}
This would be in contradiction with the present 
GGA results if we assume that the Fe atoms are incorporated
to the graphene layer as substitutionals. However, the final geometry of the 
implanted atoms in these experiments is not known. Furthermore,
a considerable amount of defects is created
during the implantation process. Although it has been argued 
that a large part of the damage is healed by vacancy-interstitial
recombination,~\cite{Barzola08} the influence of these defects,
specially interstitials and big voids, on the
observed magnetic response can be determinant. 

Therefore, it is not fully clear if we
can compare our calculations for Fe substitutional impurities
in an otherwise perfect graphene layer with these experimental data. 
However, 
as we have shown in detail in this section, 
Fe substitutionals are very 
close to a transition and, depending on
the details of the calculations, it is possible to obtain a
magnetic ground state. In particular, Fe substitutionals
develop a spin moment of 1~$\mu_B$  
at the level of GGA+U calculations for reasonable values
of the U parameter. This might be an indication that
the non-magnetic ground state found in GGA calculations
is a consequence of the limitations of the used 
functionals.

\section{Conclusions}

We have presented a DFT study of the structure, energetics, and
electronic and magnetic properties of several metal atoms
as substitutional impurities in graphene, i.e., bound to
a carbon monovacancy in the layer. We have considered 
the cases of all 3$d$ transition metals, noble metals and Zn.  
We have paid special attention to their electronic and magnetic properties
and develop a simple
model to understand the observed trends. 
Our model is based on the hybridization of 
the states of the metal atoms
with those of 
an unreconstructed carbon vacancy.
The main ingredients of the model are the assumption,
after our calculations, of 
a three-fold symmetric bonding configuration
and the approximate knowledge of
the relative energy  positions 
of the levels of the carbon monovacancy and the $d$ shell of the metal
impurity as we move along the transition series. 
With this model we can
understand the observed the variations of the 
electronic structure of the defect, the size
and localization of the spin moment, and the binding energy.
We have identified three different regimes corresponding 
to filling of carbon-metal hybridization shells with 
different character: bonding,
non-bonding and antibonding. 

In more detail: 

({\it i }) Most substitutional metal impurities present
an almost perfecly symmetric three-fold 
configuration with C$_{3v}$ symmetry. 
Noble metals slightly depart from this perfect configuration.
Only Zn
presents a considerable structural distortion. 

({\it ii}) For Sc and Ti the metal-carbon bonding shell
is completely filled. Therefore, these impurities
present the highest binding energies and zero spin moments.
Sc substitutionals act as p-dopants for graphene: 
each Sc impurity localizes one extra electron from the
carbon layer.

({\it iii}) The non-bonding $d$ shell becomes partially
populated for V, Cr and Mn, which develop a very 
localized spin-moment of 1, 2 and 3~$\mu_B$, respectively.
The binding energy decreases slightly as the $d$ shell
moves to lower energies, thus reducing its hybridization
with the higher carbon vacancy levels.

({\it iv}) For Co, Ni, the noble metals and Zn, the 
metal-carbon levels are progressively populated. This gives rise
to a oscillatory behavior of the spin moment between 0 and 1~$\mu_B$. 
The spin moments
are more delocalized than those found for V, Cr and Mn and present
a considerable contribution from the carbon atoms around the impurity.
The binding energy 
presents a local maximum for Co, but suddenly drops for 
the noble metals and has its minimum for Zn. 

({\it v}) The electronic structure, nearby E$_F$, 
of substitutional impurities of Co, Ni, the noble metals 
and Zn has a strong resemblance to that of the
unreconstructed D$_{3h}$ carbon
monovacancy 
We can draw an analogy between the electronic 
structure of these impurties and that of the
unreconstructed D$_{3h}$ carbon monovacancy 
with different number of electrons (charge states). The spin moment
of these impurities can be fully understand exploiting this equivalence.
In particular, the result that noble metals develop a spin moment
of 1~$\mu_B$ emerges naturally within this picture.

({\it vi}) For the Co impurity, the equivalence 
can be pushed a step further
and we can draw  an analogy with the electronic structure 
of a $\pi$-vacancy in a simple $\pi$-tight-binding description of
graphene. This can be used to explain the peculiar
behavior found for the magnetic couplings between Co 
substitutionals in graphene.~\cite{Santos09}

({\it vii}) Fe impurities occupy a distinct position
at the border between two different regimes. 
Their magnetic behavior stems from the
competition between the carbon-metal hybridization and
the electron-electron interaction within the 3$d$ shell.
As a result, although Fe impurities are non-magnetic
at the GGA level, GGA+U calculations with moderate
values of U (above $\sim$2~eV) produce a spin-moment of
1~$\mu_B$. 

({\it viii}) We have found that the unexpected result that
Au substitutionals~\cite{Malola09,Krasheninnikov09} present a spin moment
of 1$\mu_B$ also holds for all noble metals,
i.e., also Ag and Cu present a 1~$\mu_B$ spin moment. 

({\it ix}) 
We have found that the ground state of the Zn 
substitutional is non-magnetic due
to a Jahn-Teller distortion. 
Yet, it
is possible to stabilize a symmetric configuration with a spin moment
of 2$\mu_B$ with a very small energy penalty of $\sim$150~meV.

In summary, substitutional impurities of metals 
in graphene present some interesting magnetic and electronic
properties and, therefore, can provide an interesting route to 
add functionalities or to tune the response of devices based
on graphenic materials. 
Furthermore, recent experiments by 
Rodriguez-Manzo and Banhart~\cite{Banhart09} have 
demonstrated the possibility to create individual vacancies at
desired locations in carbon nanotubes using electron beams. 
This ability, in combination with the high stability of
substitutional impurities, can  open a route
to fabricate ordered arrays of these 
impurities at predefined locations. 
Such devices would allow, among other applications, 
the experimental
verification of the theoretical
predictions 
of unsual magnetic interactions mediated by 
graphene.~\cite{Brey07,Kirwan08,Santos09}

\appendix
\section{Pseudopententials and basis orbitals radii}
\label{sec:appendix}

For the
{\sc Siesta} calculations we have used
Troullier-Martins norm-conserving
pseudopotentials~\cite{TM} generated using the pseudization radii
shown in Table~\ref{tab:pseudos} 
of this Appendix.
The pseudopotentials for the metal atoms
include nonlinear core corrections~\cite{nlcc} for exchange
and correlation. The pseudocore
radii ($r_{core}$) have been optimized
for each element. These pseudopententials have
been tested to 
reproduce the correct results in bulk phases.

For most elements, the shape and 
radii of the basis sets of numerical 
atomic orbitals (NAOs) used with {\sc Siesta} 
were
determined automatically by the code using an
\textit{energy shift} parameter
of 50~meV.~\cite{junquera01,siesta2} However,  
for some metal atoms the binding energies where slightly
overestimated using these basis sets, mainly due to the
confinement of the free atom. 
For those atoms we enlarged the radii of the basis orbitals 
(using smaller values of the \textit{energy shift} parameter)
until binding energies were converged within a few tens of meV. The
largest radius for each element and $l$ channel 
is shown in Table~\ref{tab:orbitals}. 
Notice that double-$\zeta$ polarized (DZP) basis of these metals
include a $p$ shell. In this case the radii of $p$ orbitals is taken
equal to that of the $s$ shell.

\label{sec:append}
\begin{table} 
\caption{Cutoff radii (in Bohrs) used for the generation of the Troullier-Martins~\cite{TM}
norm-conserving
pseudopotentials used in the {\sc Siesta} calculations. $r_l$ stands for the 
cut-off radius used for the $l$ channel, while $r_{core}$ is the matching radius
used for the construction of pseudocores in order to include non-linear core corrections 
for exchange and correlation~\cite{nlcc}. }
\begin{ruledtabular}
\begin{tabular}{lccccc}
     &  Valence      &  $r_s$ ($a_o$)  & $r_p$ ($a_o$)   &  $r_d$ ($a_o$)  & $r_{core}$ \\  \hline
      & & & & & \\
Sc   &[Ar]4$s^1$3$d^2$& 3.30          & 3.30            &  1.44           &      0.89     \\
Ti   &[Ar]4$s^1$3$d^3$& 2.96          & 2.96            &  1.45           &     0.72     \\
V    &[Ar]4$s^1$3$d^4$& 2.40          & 2.79            &  1.46           &     0.69     \\
Cr   &[Ar]4$s^1$3$d^5$& 2.51          & 2.80            &  1.46           &     0.65     \\
Mn   &[Ar]4$s^1$3$d^6$& 2.51          & 2.77            &  1.45           &      0.60     \\
Fe   &[Ar]4$s^1$3$d^7$& 2.10          & 2.10            &  1.68           &      0.67     \\
Co   &[Ar]4$s^2$3$d^7$& 2.37          & 2.48            &  1.68           &     0.67      \\
Ni   &[Ar]4$s^2$3$d^8$& 1.85          &   1.95        &   1.45            &      0.53       \\
Cu   &[Ar]4$s^1$3$d^{10}$& 2.33          &   2.30      &   1.79           &    0.53        \\
Ag   &[Kr]5$s^1$4$d^{10}$&  2.45         &  2.58       & 2.00              &   0.83          \\
Au   &[Xe,4$f^{14}$]6$s^1$5$d^{10}$&  2.55   &  2.68  & 2.20             &    0.93      \\
Zn   &[Ar]4$s^2$3$d^{10}$& 2.04      &  2.21      &   1.66         &      0.49      \\
C    &[He]2$s^2$2$p^2$  &  1.25 & 1.25 & 1.25 &  --  \\
\end{tabular}
\end{ruledtabular}
\label{tab:pseudos}
\end{table}

\begin{table}
\caption{Cutoff radii (in Bohrs)  of the numerical atomic 
orbitals (NAOs) used in {\sc Siesta} as a basis set.
Radii used for the $s$ and polarization $p$ shell
are equal. }
\begin{ruledtabular}
\begin{tabular}{lcc}
     &  $r^{NAO}_{sp}$ ($a_o$)  &   $r^{NAO}_d$ ($a_o$) \\  \hline
     &                       &                    \\
Co   &  8.00  & 4.73  \\
Ni   &  10.94  & 6.81  \\
Cu   &  8.87  &  5.52  \\
Ag   &  10.48 &  6.52   \\
Au   &  8.63  &  6.08 \\
Zn   &  9.24  &  5.33  \\
\end{tabular}
\end{ruledtabular}
\label{tab:orbitals}
\end{table}

\begin{acknowledgments}
We acknowledge support from Basque
Departamento de
Educaci\'on
and the UPV/EHU (Grant
No. IT-366-07), the Spanish Ministerio de Educaci\'on
y Ciencia (Grant No. FIS2007-66711-C02-02) and the
ETORTEK program funded by the Basque Departamento de 
Industria and the Diputaci\'on Foral de Guipuzcoa. 
EJGS would like to thank N. Gonz\'alez-Lakunza for 
valuable help with 
the VASP code.
\end{acknowledgments}


\begin{thebibliography}{47}
\expandafter\ifx\csname natexlab\endcsname\relax\def\natexlab#1{#1}\fi
\expandafter\ifx\csname bibnamefont\endcsname\relax
  \def\bibnamefont#1{#1}\fi
\expandafter\ifx\csname bibfnamefont\endcsname\relax
  \def\bibfnamefont#1{#1}\fi
\expandafter\ifx\csname citenamefont\endcsname\relax
  \def\citenamefont#1{#1}\fi
\expandafter\ifx\csname url\endcsname\relax
  \def\url#1{\texttt{#1}}\fi
\expandafter\ifx\csname urlprefix\endcsname\relax\def\urlprefix{URL }\fi
\providecommand{\bibinfo}[2]{#2}
\providecommand{\eprint}[2][]{\url{#2}}

\bibitem[{\citenamefont{Geim and Novoselov}(2007)}]{Geim07}
\bibinfo{author}{\bibfnamefont{A.~K.} \bibnamefont{Geim}} \bibnamefont{and}
  \bibinfo{author}{\bibfnamefont{K.~S.} \bibnamefont{Novoselov}},
  \bibinfo{journal}{Nature Mat.} \textbf{\bibinfo{volume}{6}},
  \bibinfo{pages}{183} (\bibinfo{year}{2007}).

\bibitem[{\citenamefont{{Castro Neto} et~al.}(2009)\citenamefont{{Castro Neto},
  Guinea, Peres, Novoselov, and Geim}}]{Neto09}
\bibinfo{author}{\bibfnamefont{A.~H.} \bibnamefont{{Castro Neto}}},
  \bibinfo{author}{\bibfnamefont{F.}~\bibnamefont{Guinea}},
  \bibinfo{author}{\bibfnamefont{N.~M.} \bibnamefont{Peres}},
  \bibinfo{author}{\bibfnamefont{K.~S.} \bibnamefont{Novoselov}},
  \bibnamefont{and} \bibinfo{author}{\bibfnamefont{A.~K.} \bibnamefont{Geim}},
  \bibinfo{journal}{Rev. Mod. Phys.} \textbf{\bibinfo{volume}{81}},
  \bibinfo{pages}{109} (\bibinfo{year}{2009}).

\bibitem[{\citenamefont{Novoselov et~al.}(2005)}]{Novoselov05}
\bibinfo{author}{\bibfnamefont{K.~S.} \bibnamefont{Novoselov}}
  \bibnamefont{et~al.}, \bibinfo{journal}{Nature (London)}
  \textbf{\bibinfo{volume}{438}}, \bibinfo{pages}{197} (\bibinfo{year}{2005}).

\bibitem[{\citenamefont{Katsnelson et~al.}(2006)\citenamefont{Katsnelson,
  Novoselov, and Geim}}]{Katsnelson06}
\bibinfo{author}{\bibfnamefont{M.~I.} \bibnamefont{Katsnelson}},
  \bibinfo{author}{\bibfnamefont{K.~S.} \bibnamefont{Novoselov}},
  \bibnamefont{and} \bibinfo{author}{\bibfnamefont{A.~K.} \bibnamefont{Geim}},
  \bibinfo{journal}{Nature Physics} \textbf{\bibinfo{volume}{2}},
  \bibinfo{pages}{620} (\bibinfo{year}{2006}).

\bibitem[{\citenamefont{Park et~al.}(2008)\citenamefont{Park, Son, Cohen, and
  Louie}}]{Park08}
\bibinfo{author}{\bibfnamefont{C.}~\bibnamefont{Park}},
  \bibinfo{author}{\bibfnamefont{Y.-W.} \bibnamefont{Son}},
  \bibinfo{author}{\bibfnamefont{M.~L.} \bibnamefont{Cohen}}, \bibnamefont{and}
  \bibinfo{author}{\bibfnamefont{S.~G.} \bibnamefont{Louie}},
  \bibinfo{journal}{Nature Physics} \textbf{\bibinfo{volume}{4}},
  \bibinfo{pages}{213} (\bibinfo{year}{2008}).

\bibitem[{\citenamefont{Avouris et~al.}(2007)\citenamefont{Avouris, Chen, and
  Perebeinos}}]{Avouris07}
\bibinfo{author}{\bibfnamefont{P.}~\bibnamefont{Avouris}},
  \bibinfo{author}{\bibfnamefont{Z.}~\bibnamefont{Chen}}, \bibnamefont{and}
  \bibinfo{author}{\bibfnamefont{V.}~\bibnamefont{Perebeinos}},
  \bibinfo{journal}{Nature Nanotechnology} \textbf{\bibinfo{volume}{2}},
  \bibinfo{pages}{605} (\bibinfo{year}{2007}).

\bibitem[{\citenamefont{Trauzettel et~al.}(2007)}]{Trauzettel07}
\bibinfo{author}{\bibfnamefont{B.}~\bibnamefont{Trauzettel}}
  \bibnamefont{et~al.}, \bibinfo{journal}{Nature Phys.}
  \textbf{\bibinfo{volume}{3}}, \bibinfo{pages}{192} (\bibinfo{year}{2007}).

\bibitem[{\citenamefont{Son et~al.}(2006)\citenamefont{Son, Cohen, and
  Louie}}]{Son06}
\bibinfo{author}{\bibfnamefont{Y.-W.} \bibnamefont{Son}},
  \bibinfo{author}{\bibfnamefont{M.~L.} \bibnamefont{Cohen}}, \bibnamefont{and}
  \bibinfo{author}{\bibfnamefont{S.~G.} \bibnamefont{Louie}},
  \bibinfo{journal}{Nature} \textbf{\bibinfo{volume}{444}},
  \bibinfo{pages}{347} (\bibinfo{year}{2006}).

\bibitem[{\citenamefont{Yazyev and Katsnelson}(2008)}]{Yazyev08}
\bibinfo{author}{\bibfnamefont{O.~V.} \bibnamefont{Yazyev}} \bibnamefont{and}
  \bibinfo{author}{\bibfnamefont{M.~I.} \bibnamefont{Katsnelson}},
  \bibinfo{journal}{Phys. Rev. Lett.} \textbf{\bibinfo{volume}{100}},
  \bibinfo{pages}{047209} (\bibinfo{year}{2008}).

\bibitem[{\citenamefont{Hueso et~al.}(2007)}]{Hueso07}
\bibinfo{author}{\bibfnamefont{L.~E.} \bibnamefont{Hueso}}
  \bibnamefont{et~al.}, \bibinfo{journal}{Nature}
  \textbf{\bibinfo{volume}{445}}, \bibinfo{pages}{410} (\bibinfo{year}{2007}).

\bibitem[{\citenamefont{Esquinazi et~al.}(2003)\citenamefont{Esquinazi,
  Spemann, H{\"o}hne, Setzer, Han, and Butz}}]{Esquinazi03}
\bibinfo{author}{\bibfnamefont{P.}~\bibnamefont{Esquinazi}},
  \bibinfo{author}{\bibfnamefont{D.}~\bibnamefont{Spemann}},
  \bibinfo{author}{\bibfnamefont{D.}~\bibnamefont{H{\"o}hne}},
  \bibinfo{author}{\bibfnamefont{A.}~\bibnamefont{Setzer}},
  \bibinfo{author}{\bibfnamefont{K.-H.} \bibnamefont{Han}}, \bibnamefont{and}
  \bibinfo{author}{\bibfnamefont{T.}~\bibnamefont{Butz}},
  \bibinfo{journal}{Phys. Rev. Lett.} \textbf{\bibinfo{volume}{91}},
  \bibinfo{pages}{227201} (\bibinfo{year}{2003}).

\bibitem[{\citenamefont{Lehtinen et~al.}(2003)\citenamefont{Lehtinen, Foster,
  Ayuela, Krasheninnikov, Nordlund, and Nieminen}}]{Lehtinen03}
\bibinfo{author}{\bibfnamefont{P.~O.} \bibnamefont{Lehtinen}},
  \bibinfo{author}{\bibfnamefont{A.~S.} \bibnamefont{Foster}},
  \bibinfo{author}{\bibfnamefont{A.}~\bibnamefont{Ayuela}},
  \bibinfo{author}{\bibfnamefont{A.~V.} \bibnamefont{Krasheninnikov}},
  \bibinfo{author}{\bibfnamefont{K.}~\bibnamefont{Nordlund}}, \bibnamefont{and}
  \bibinfo{author}{\bibfnamefont{R.}~\bibnamefont{Nieminen}},
  \bibinfo{journal}{Phys. Rev. Lett.} \textbf{\bibinfo{volume}{91}},
  \bibinfo{pages}{017202} (\bibinfo{year}{2003}).

\bibitem[{\citenamefont{Lehtinen et~al.}(2004)\citenamefont{Lehtinen, Foster,
  Ma, Krasheninnikov, and Nieminen}}]{Lehtinen04}
\bibinfo{author}{\bibfnamefont{P.~O.} \bibnamefont{Lehtinen}},
  \bibinfo{author}{\bibfnamefont{A.~S.} \bibnamefont{Foster}},
  \bibinfo{author}{\bibfnamefont{Y.}~\bibnamefont{Ma}},
  \bibinfo{author}{\bibfnamefont{A.~V.} \bibnamefont{Krasheninnikov}},
  \bibnamefont{and} \bibinfo{author}{\bibfnamefont{R.}~\bibnamefont{Nieminen}},
  \bibinfo{journal}{Phys. Rev. Lett.} \textbf{\bibinfo{volume}{93}},
  \bibinfo{pages}{187202} (\bibinfo{year}{2004}).

\bibitem[{\citenamefont{Makarova and Palacios}(2006)}]{Palacios06}
\bibinfo{editor}{\bibfnamefont{T.}~\bibnamefont{Makarova}} \bibnamefont{and}
  \bibinfo{editor}{\bibfnamefont{F.}~\bibnamefont{Palacios}}, eds.,
  \emph{\bibinfo{title}{Carbon Based Magnetism}} (\bibinfo{publisher}{Elsevier,
  Amsterdam}, \bibinfo{year}{2006}).

\bibitem[{\citenamefont{Krashninnikov and Banhart}(2007)}]{Krasheninnikov07}
\bibinfo{author}{\bibfnamefont{A.~V.} \bibnamefont{Krashninnikov}}
  \bibnamefont{and} \bibinfo{author}{\bibfnamefont{F.}~\bibnamefont{Banhart}},
  \bibinfo{journal}{Nature Mat.} \textbf{\bibinfo{volume}{6}},
  \bibinfo{pages}{723} (\bibinfo{year}{2007}).

\bibitem[{\citenamefont{Uchoa et~al.}(2008)}]{Uchoa08}
\bibinfo{author}{\bibfnamefont{B.}~\bibnamefont{Uchoa}} \bibnamefont{et~al.},
  \bibinfo{journal}{Phys. Rev. Lett.} \textbf{\bibinfo{volume}{101}},
  \bibinfo{pages}{026805} (\bibinfo{year}{2008}).

\bibitem[{\citenamefont{Yazyev}(2008)}]{Yazyev08bis}
\bibinfo{author}{\bibfnamefont{O.~V.} \bibnamefont{Yazyev}},
  \bibinfo{journal}{Phys. Rev. Lett.} \textbf{\bibinfo{volume}{101}},
  \bibinfo{pages}{037203} (\bibinfo{year}{2008}).

\bibitem[{\citenamefont{J.~J.~Palacios and Brey}(2008)}]{Palacios08}
\bibinfo{author}{\bibfnamefont{J.~F.-R.} \bibnamefont{J.~J.~Palacios}}
  \bibnamefont{and} \bibinfo{author}{\bibfnamefont{L.}~\bibnamefont{Brey}},
  \bibinfo{journal}{Phys. Rev. B} \textbf{\bibinfo{volume}{77}},
  \bibinfo{pages}{195428} (\bibinfo{year}{2008}).

\bibitem[{\citenamefont{Fern{\'a}ndez-Rossier and Palacios}(2007)}]{Rossier07}
\bibinfo{author}{\bibfnamefont{J.}~\bibnamefont{Fern{\'a}ndez-Rossier}}
  \bibnamefont{and} \bibinfo{author}{\bibfnamefont{J.~J.}
  \bibnamefont{Palacios}}, \bibinfo{journal}{Phys. Rev. Lett.}
  \textbf{\bibinfo{volume}{99}}, \bibinfo{pages}{177204}
  (\bibinfo{year}{2007}).

\bibitem[{\citenamefont{Gan et~al.}(2008)\citenamefont{Gan, Sun, and
  Banhart}}]{Banhart08}
\bibinfo{author}{\bibfnamefont{Y.}~\bibnamefont{Gan}},
  \bibinfo{author}{\bibfnamefont{L.}~\bibnamefont{Sun}}, \bibnamefont{and}
  \bibinfo{author}{\bibfnamefont{F.}~\bibnamefont{Banhart}},
  \bibinfo{journal}{Small} \textbf{\bibinfo{volume}{4}}, \bibinfo{pages}{587}
  (\bibinfo{year}{2008}).

\bibitem[{\citenamefont{Ushiro et~al.}(2006)\citenamefont{Ushiro, Uno,
  Fujikawa, Sato, Tohji, Watari, Chun, Koike, and Asakura}}]{Ushiro06}
\bibinfo{author}{\bibfnamefont{M.}~\bibnamefont{Ushiro}},
  \bibinfo{author}{\bibfnamefont{K.}~\bibnamefont{Uno}},
  \bibinfo{author}{\bibfnamefont{T.}~\bibnamefont{Fujikawa}},
  \bibinfo{author}{\bibfnamefont{Y.}~\bibnamefont{Sato}},
  \bibinfo{author}{\bibfnamefont{K.}~\bibnamefont{Tohji}},
  \bibinfo{author}{\bibfnamefont{F.}~\bibnamefont{Watari}},
  \bibinfo{author}{\bibfnamefont{W.~J.} \bibnamefont{Chun}},
  \bibinfo{author}{\bibfnamefont{Y.}~\bibnamefont{Koike}}, \bibnamefont{and}
  \bibinfo{author}{\bibfnamefont{K.}~\bibnamefont{Asakura}},
  \bibinfo{journal}{Phys. Rev. B} \textbf{\bibinfo{volume}{73}},
  \bibinfo{pages}{144103} (\bibinfo{year}{2006}).

\bibitem[{\citenamefont{Santos et~al.}(2008)\citenamefont{Santos, Ayuela,
  Fagan, andD. L.~Azevedo, Filho, and S\'anchez-Portal}}]{Santos08}
\bibinfo{author}{\bibfnamefont{E.~J.~G.} \bibnamefont{Santos}},
  \bibinfo{author}{\bibfnamefont{A.}~\bibnamefont{Ayuela}},
  \bibinfo{author}{\bibfnamefont{S.~B.} \bibnamefont{Fagan}},
  \bibinfo{author}{\bibfnamefont{J.~M.~F.} \bibnamefont{andD. L.~Azevedo}},
  \bibinfo{author}{\bibfnamefont{A.~G.~S.} \bibnamefont{Filho}},
  \bibnamefont{and}
  \bibinfo{author}{\bibfnamefont{D.}~\bibnamefont{S\'anchez-Portal}},
  \bibinfo{journal}{Phys. Rev. B} \textbf{\bibinfo{volume}{78}},
  \bibinfo{pages}{195420} (\bibinfo{year}{2008}).

\bibitem[{\citenamefont{Banhart et~al.}(2000)\citenamefont{Banhart, Charlier,
  and Ajayan}}]{Banhart00}
\bibinfo{author}{\bibfnamefont{F.}~\bibnamefont{Banhart}},
  \bibinfo{author}{\bibfnamefont{J.~C.} \bibnamefont{Charlier}},
  \bibnamefont{and} \bibinfo{author}{\bibfnamefont{P.~M.}
  \bibnamefont{Ajayan}}, \bibinfo{journal}{Phys. Rev. Lett.}
  \textbf{\bibinfo{volume}{84}}, \bibinfo{pages}{686} (\bibinfo{year}{2000}).

\bibitem[{\citenamefont{Dresselhaus et~al.}(2001)\citenamefont{Dresselhaus,
  Dresselhaus, and Avouris}}]{growth}
\bibinfo{editor}{\bibfnamefont{M.~S.} \bibnamefont{Dresselhaus}},
  \bibinfo{editor}{\bibfnamefont{G.}~\bibnamefont{Dresselhaus}},
  \bibnamefont{and} \bibinfo{editor}{\bibfnamefont{P.}~\bibnamefont{Avouris}},
  eds., \emph{\bibinfo{title}{Carbon Nanotubes: Synthesis, Structure,
  Properties and Applications}} (\bibinfo{publisher}{Springer, Berlin},
  \bibinfo{year}{2001}).

\bibitem[{\citenamefont{Rodriguez-Manzo and Banhart}(2009)}]{Banhart09}
\bibinfo{author}{\bibfnamefont{J.~A.} \bibnamefont{Rodriguez-Manzo}}
  \bibnamefont{and} \bibinfo{author}{\bibfnamefont{F.}~\bibnamefont{Banhart}},
  \bibinfo{journal}{Nano Letters} \textbf{\bibinfo{volume}{9}},
  \bibinfo{pages}{2285} (\bibinfo{year}{2009}).

\bibitem[{\citenamefont{Brey et~al.}(2007)\citenamefont{Brey, Ferting, and
  Sarma}}]{Brey07}
\bibinfo{author}{\bibfnamefont{L.}~\bibnamefont{Brey}},
  \bibinfo{author}{\bibfnamefont{H.~A.} \bibnamefont{Ferting}},
  \bibnamefont{and} \bibinfo{author}{\bibfnamefont{S.~D.} \bibnamefont{Sarma}},
  \bibinfo{journal}{Phys. Rev. Lett.} \textbf{\bibinfo{volume}{99}},
  \bibinfo{pages}{116802} (\bibinfo{year}{2007}).

\bibitem[{\citenamefont{Kirwan et~al.}(2008)}]{Kirwan08}
\bibinfo{author}{\bibfnamefont{D.~F.} \bibnamefont{Kirwan}}
  \bibnamefont{et~al.}, \bibinfo{journal}{Phys. Rev. B}
  \textbf{\bibinfo{volume}{77}}, \bibinfo{pages}{085432}
  (\bibinfo{year}{2008}).

\bibitem[{\citenamefont{Santos et~al.}(2009)\citenamefont{Santos,
  S\'anchez-Portal, and Ayuela}}]{Santos09}
\bibinfo{author}{\bibfnamefont{E.~J.~G.} \bibnamefont{Santos}},
  \bibinfo{author}{\bibfnamefont{D.}~\bibnamefont{S\'anchez-Portal}},
  \bibnamefont{and} \bibinfo{author}{\bibfnamefont{A.}~\bibnamefont{Ayuela}}
  (\bibinfo{year}{2009}), \eprint{cond-mat/0906.5604}.

\bibitem[{\citenamefont{Krasheninnikov
  et~al.}(2009)\citenamefont{Krasheninnikov, Lehtinen, Foster, Pyykk{\"o}, and
  Nieminen}}]{Krasheninnikov09}
\bibinfo{author}{\bibfnamefont{A.~V.} \bibnamefont{Krasheninnikov}},
  \bibinfo{author}{\bibfnamefont{P.~O.} \bibnamefont{Lehtinen}},
  \bibinfo{author}{\bibfnamefont{A.~S.} \bibnamefont{Foster}},
  \bibinfo{author}{\bibfnamefont{P.}~\bibnamefont{Pyykk{\"o}}},
  \bibnamefont{and} \bibinfo{author}{\bibfnamefont{R.~M.}
  \bibnamefont{Nieminen}}, \bibinfo{journal}{Phys. Rev. Lett.}
  \textbf{\bibinfo{volume}{102}}, \bibinfo{pages}{126807}
  (\bibinfo{year}{2009}).

\bibitem[{\citenamefont{Malola et~al.}(2009)\citenamefont{Malola, H{\"a}kkinen,
  and Koskinen}}]{Malola09}
\bibinfo{author}{\bibfnamefont{S.}~\bibnamefont{Malola}},
  \bibinfo{author}{\bibfnamefont{H.}~\bibnamefont{H{\"a}kkinen}},
  \bibnamefont{and} \bibinfo{author}{\bibfnamefont{P.}~\bibnamefont{Koskinen}},
  \bibinfo{journal}{Appl. Phys. Lett.} \textbf{\bibinfo{volume}{94}},
  \bibinfo{pages}{043106} (\bibinfo{year}{2009}).

\bibitem[{\citenamefont{Boukhalov and Katsnelson}(2009)}]{Boukhvalov09}
\bibinfo{author}{\bibfnamefont{D.~W.} \bibnamefont{Boukhalov}}
  \bibnamefont{and} \bibinfo{author}{\bibfnamefont{M.~I.}
  \bibnamefont{Katsnelson}}, \bibinfo{journal}{Appl. Phys. Lett.}
  \textbf{\bibinfo{volume}{95}}, \bibinfo{pages}{023109}
  (\bibinfo{year}{2009}).

\bibitem[{\citenamefont{Suarez-Martinez and Ewels}(2009)}]{Ewels09}
\bibinfo{author}{\bibfnamefont{I.}~\bibnamefont{Suarez-Martinez}}
  \bibnamefont{and} \bibinfo{author}{\bibfnamefont{C.~P.} \bibnamefont{Ewels}}
  (\bibinfo{year}{2009}), \bibinfo{note}{private communication}.

\bibitem[{\citenamefont{Perdew et~al.}(1996)\citenamefont{Perdew, Burke, and
  Ernzerhof}}]{gga}
\bibinfo{author}{\bibfnamefont{J.~P.} \bibnamefont{Perdew}},
  \bibinfo{author}{\bibfnamefont{K.}~\bibnamefont{Burke}}, \bibnamefont{and}
  \bibinfo{author}{\bibfnamefont{M.}~\bibnamefont{Ernzerhof}},
  \bibinfo{journal}{Phys. Rev. Lett.} \textbf{\bibinfo{volume}{77}},
  \bibinfo{pages}{3865} (\bibinfo{year}{1996}).

\bibitem[{\citenamefont{S\'anchez-Portal
  et~al.}(1997)\citenamefont{S\'anchez-Portal, Artacho, and Soler}}]{siesta1}
\bibinfo{author}{\bibfnamefont{D.}~\bibnamefont{S\'anchez-Portal}},
  \bibinfo{author}{\bibfnamefont{P.~O.~E.} \bibnamefont{Artacho}},
  \bibnamefont{and} \bibinfo{author}{\bibfnamefont{J.~M.} \bibnamefont{Soler}},
  \bibinfo{journal}{Int. J. Quantum Chem.} \textbf{\bibinfo{volume}{65}},
  \bibinfo{pages}{453} (\bibinfo{year}{1997}).

\bibitem[{\citenamefont{Soler et~al.}(2002)\citenamefont{Soler, Artacho, Gale,
  Garc{\'i}a, Junquera, Ordej\'on, and S\'anchez-Portal}}]{siesta2}
\bibinfo{author}{\bibfnamefont{J.~M.} \bibnamefont{Soler}},
  \bibinfo{author}{\bibfnamefont{E.}~\bibnamefont{Artacho}},
  \bibinfo{author}{\bibfnamefont{J.~D.} \bibnamefont{Gale}},
  \bibinfo{author}{\bibfnamefont{A.}~\bibnamefont{Garc{\'i}a}},
  \bibinfo{author}{\bibfnamefont{J.}~\bibnamefont{Junquera}},
  \bibinfo{author}{\bibfnamefont{P.}~\bibnamefont{Ordej\'on}},
  \bibnamefont{and}
  \bibinfo{author}{\bibfnamefont{D.}~\bibnamefont{S\'anchez-Portal}},
  \bibinfo{journal}{J. Phys.: Condensed Matter} \textbf{\bibinfo{volume}{14}},
  \bibinfo{pages}{2745} (\bibinfo{year}{2002}).

\bibitem[{\citenamefont{S\'anchez-Portal
  et~al.}(2004)\citenamefont{S\'anchez-Portal, Ordej\'on, and
  Canadell}}]{siesta3}
\bibinfo{author}{\bibfnamefont{D.}~\bibnamefont{S\'anchez-Portal}},
  \bibinfo{author}{\bibfnamefont{P.}~\bibnamefont{Ordej\'on}},
  \bibnamefont{and} \bibinfo{author}{\bibfnamefont{E.}~\bibnamefont{Canadell}},
  \bibinfo{journal}{Structure and Bonding} \textbf{\bibinfo{volume}{113}},
  \bibinfo{pages}{103} (\bibinfo{year}{2004}).

\bibitem[{\citenamefont{Junquera et~al.}(2001)\citenamefont{Junquera, Paz,
  S\'anchez-Portal, and Artacho}}]{junquera01}
\bibinfo{author}{\bibfnamefont{J.}~\bibnamefont{Junquera}},
  \bibinfo{author}{\bibfnamefont{O.}~\bibnamefont{Paz}},
  \bibinfo{author}{\bibfnamefont{D.}~\bibnamefont{S\'anchez-Portal}},
  \bibnamefont{and} \bibinfo{author}{\bibfnamefont{E.}~\bibnamefont{Artacho}},
  \bibinfo{journal}{Phys. Rev. B} \textbf{\bibinfo{volume}{64}},
  \bibinfo{pages}{235111} (\bibinfo{year}{2001}).

\bibitem[{\citenamefont{Kresse and Hafner}(1993)}]{vasp1}
\bibinfo{author}{\bibfnamefont{G.}~\bibnamefont{Kresse}} \bibnamefont{and}
  \bibinfo{author}{\bibfnamefont{J.}~\bibnamefont{Hafner}},
  \bibinfo{journal}{Phys. Rev. B} \textbf{\bibinfo{volume}{47}},
  \bibinfo{pages}{558} (\bibinfo{year}{1993}).

\bibitem[{\citenamefont{Kresse and Furthm{\"u}ller}(1996)}]{vasp2}
\bibinfo{author}{\bibfnamefont{G.}~\bibnamefont{Kresse}} \bibnamefont{and}
  \bibinfo{author}{\bibfnamefont{J.}~\bibnamefont{Furthm{\"u}ller}},
  \bibinfo{journal}{Phys. Rev. B} \textbf{\bibinfo{volume}{54}},
  \bibinfo{pages}{11169} (\bibinfo{year}{1996}).

\bibitem[{\citenamefont{Monkhosrt and Pack}(1976)}]{MonkhorstPack}
\bibinfo{author}{\bibfnamefont{H.~J.} \bibnamefont{Monkhosrt}}
  \bibnamefont{and} \bibinfo{author}{\bibfnamefont{J.~D.} \bibnamefont{Pack}},
  \bibinfo{journal}{Phys. Rev. B} \textbf{\bibinfo{volume}{13}},
  \bibinfo{pages}{5188} (\bibinfo{year}{1976}).

\bibitem[{\citenamefont{Troullier and Martins}(1991)}]{TM}
\bibinfo{author}{\bibfnamefont{N.}~\bibnamefont{Troullier}} \bibnamefont{and}
  \bibinfo{author}{\bibfnamefont{J.~L.} \bibnamefont{Martins}},
  \bibinfo{journal}{Phys. Rev. B} \textbf{\bibinfo{volume}{43}},
  \bibinfo{pages}{1993} (\bibinfo{year}{1991}).

\bibitem[{\citenamefont{Louie et~al.}(1982)\citenamefont{Louie, Froyen, and
  Cohen}}]{nlcc}
\bibinfo{author}{\bibfnamefont{S.~G.} \bibnamefont{Louie}},
  \bibinfo{author}{\bibfnamefont{S.}~\bibnamefont{Froyen}}, \bibnamefont{and}
  \bibinfo{author}{\bibfnamefont{M.~L.} \bibnamefont{Cohen}},
  \bibinfo{journal}{Phys. Rev. B} \textbf{\bibinfo{volume}{26}},
  \bibinfo{pages}{1738} (\bibinfo{year}{1982}).

\bibitem[{\citenamefont{Dudarev et~al.}(1998)\citenamefont{Dudarev, Botton,
  Savrasov, Humpreys, and Sutton}}]{Dudarev98}
\bibinfo{author}{\bibfnamefont{S.~L.} \bibnamefont{Dudarev}},
  \bibinfo{author}{\bibfnamefont{G.~A.} \bibnamefont{Botton}},
  \bibinfo{author}{\bibfnamefont{S.~Y.} \bibnamefont{Savrasov}},
  \bibinfo{author}{\bibfnamefont{C.~J.} \bibnamefont{Humpreys}},
  \bibnamefont{and} \bibinfo{author}{\bibfnamefont{A.~P.}
  \bibnamefont{Sutton}}, \bibinfo{journal}{Phys. Rev. B}
  \textbf{\bibinfo{volume}{57}}, \bibinfo{pages}{1505} (\bibinfo{year}{1998}).

\bibitem[{\citenamefont{S\'anchez-Portal
  et~al.}(2009)\citenamefont{S\'anchez-Portal, Santos, and Ayuela}}]{Portal09}
\bibinfo{author}{\bibfnamefont{D.}~\bibnamefont{S\'anchez-Portal}},
  \bibinfo{author}{\bibfnamefont{E.~J.~G.} \bibnamefont{Santos}},
  \bibnamefont{and} \bibinfo{author}{\bibfnamefont{A.}~\bibnamefont{Ayuela}}
  (\bibinfo{year}{2009}), \bibinfo{note}{submitted}.

\bibitem[{nom()}]{nomenclature}
\bibinfo{note}{For simplicity and consistency we use throughout the paper the
  same nomenclature to label the different defect levels. However, it should be
  understood that for the noble metals and Zn the so-called antibonding A
  $p_z$-$d_{z^2}$ level presents a sizeable contribution from the metal $s$
  orbital, while the degenerate E $sp$-$d$ levels have non-negligible
  contributions from the $p$ orbitals of the metal.}

\bibitem[{\citenamefont{Sielemann et~al.}(2008)\citenamefont{Sielemann,
  Kobayashi, Yoshida, Gunnlaugsson, and Weyer}}]{Sielemann08}
\bibinfo{author}{\bibfnamefont{R.}~\bibnamefont{Sielemann}},
  \bibinfo{author}{\bibfnamefont{Y.}~\bibnamefont{Kobayashi}},
  \bibinfo{author}{\bibfnamefont{Y.}~\bibnamefont{Yoshida}},
  \bibinfo{author}{\bibfnamefont{H.~P.} \bibnamefont{Gunnlaugsson}},
  \bibnamefont{and} \bibinfo{author}{\bibfnamefont{G.}~\bibnamefont{Weyer}},
  \bibinfo{journal}{Phys. Rev. Lett.} \textbf{\bibinfo{volume}{101}},
  \bibinfo{pages}{137206} (\bibinfo{year}{2008}).

\bibitem[{\citenamefont{Barzola-Quiquia
  et~al.}(2008)\citenamefont{Barzola-Quiquia, H{\"o}hne, Rothermel, Setzer,
  Esquinazi, and Heera}}]{Barzola08}
\bibinfo{author}{\bibfnamefont{J.}~\bibnamefont{Barzola-Quiquia}},
  \bibinfo{author}{\bibfnamefont{R.}~\bibnamefont{H{\"o}hne}},
  \bibinfo{author}{\bibfnamefont{M.}~\bibnamefont{Rothermel}},
  \bibinfo{author}{\bibfnamefont{A.}~\bibnamefont{Setzer}},
  \bibinfo{author}{\bibfnamefont{P.}~\bibnamefont{Esquinazi}},
  \bibnamefont{and} \bibinfo{author}{\bibfnamefont{V.}~\bibnamefont{Heera}},
  \bibinfo{journal}{Eur. Phys. J B} \textbf{\bibinfo{volume}{61}},
  \bibinfo{pages}{127} (\bibinfo{year}{2008}).

\end{thebibliography}
\end{document}